# Recent Progress in III-V based ferromagnetic semiconductors: Band structure, Fermi level, and tunneling transport

Masaaki Tanaka[1,2], Shinobu Ohya[1], and Pham Nam Hai[1]

1.  *Department of Electrical Engineering and Information Systems, The University of Tokyo, 7-3-1 Hongo, Bunkyo-ku, Tokyo 113-8656, Japan*

2.  *Institute for Nano Quantum Electronics, The University of Tokyo, 4-6-1 Komaba, Meguro-ku, Tokyo 153-8505, Japan*

Spin-based electronics or spintronics is an emerging field, in which we try to utilize spin degrees of freedom as well as charge transport in materials and devices.   While metal-based spin-devices, such as magnetic-field sensors and magnetoresistive random access memory using giant magnetoresistance and tunneling magnetoresistance, are already put to practical use, semiconductor-based spintronics has greater potential for expansion because of good compatibility with existing semiconductor technology.   Many semiconductor-based spintronics devices with useful functionalities have been proposed and explored so far.   To realize those devices and functionalities, we definitely need appropriate materials which have both the properties of semiconductors and ferromagnets. Ferromagnetic semiconductors (FMS), which are alloy semiconductors containing magnetic atoms such as Mn and Fe, are one of the most promising classes of materials for this purpose, and thus have been intensively studied for the past two decades.   Here, we review the recent progress in the studies of the most prototypical III-V based FMS, p-type (GaMn)As, and its heterostructures with focus on tunneling transport, Fermi level, and bandstructure.   Furthermore, we cover the properties of a new n-type FMS, (InFe)As, which shows electron-induced ferromagnetism. These FMS materials having zinc-blende crystal structure show excellent compatibility with well-developed III-V heterostructures and devices.

Authors' email addresses
Masaaki Tanaka (corresponding author): masaaki@ee.t.u-tokyo.ac.jp
Shinobu Ohya: ohya@cryst.t.u-tokyo.ac.jp
Pham Nam Hai: pham@cryst.t.u-tokyo.ac.jp

TABLE OF CONTENTS

## I. INTRODUCTION

For more than two decades, the study of spin-based electronics or so-called spintronics has received much attention[1,2]. Spintronics is an emerging field, in which we try to utilize not only charge transport of carriers but also spin degrees of freedom in materials and devices. Metal-based devices, using giant magnetoresistance (GMR)[3,4] and tunneling magnetoresistance (TMR)[5,6,7], are already applied to magnetic-field sensors in hard disk drive systems, greatly contributing to the increase in data storage capacity. The GMR and TMR effects are caused by the spin transport between two ferromagnetic metal (FM) electrodes separated by an ultrathin nonmagnetic metal and an insulator tunnel barrier (I), respectively. The latter device structure consisting of FM/I/FM is called magnetic tunnel junction (MTJ), and spin tunneling in MTJ is one of the central research subjects in metal-based spintronics, because properly designed MTJs exhibiting high TMR ratios are used not only for magnetic-field sensors, but also for nonvolatile magnetoresistive random access memory (MRAM).

Introducing such spin-related properties in *semiconductors* is expected to give new spin degrees of freedom in semiconductor devices and electronics[8]. Many semiconductor-based devices with useful spin-related functionalities have been proposed and explored so far[9,10,11]. To realize such spintronics devices and functionalities, we definitely need appropriate materials which have both the properties of semiconductors and ferromagnets. The most intensively studied materials are *ferromagnetic semiconductors* (FMS), which are alloy semiconductors containing transition-metal atoms like Mn and Fe.

Early studies of ferromagnetic semiconductors such as Eu chalcogenides were carried out since the late 1960's – early 1970's and opened a research field of interplay between ferromagnetism and semiconductor properties[12].

In the 1980s, diluted magnetic semiconductors (DMS), for example (CdMn)Te and (ZnMn)Se, are fabricated by alloying II-VI semiconductors with magnetic atoms like Mn[13]. Not only bulk-like materials, but also quantum wells (QWs) and superlattices containing ultrathin DMS layers are grown by molecular beam epitaxy (MBE). II-VI based DMSs show many interesting phenomena such as large magneto-optical effects. Carrier-induced ferromagnetism was first discovered in IV-VI based magnetic semiconductors such as PbSnMnTe[14]. However, although these magnetic semiconductors show a variety of physically intriguing phenomena, they are not very compatible with modern electronic and optical devices, in which group IV and III-V semiconductors are mainly used.

Successful growth of Mn doped III-V semiconductors, (InMn)As[15,16,17,18] and (GaMn)As[19,20,21], by low-temperature molecular-beam epitaxy (LT-MBE) and discovery of carrier-induced ferromagnetism have led to a new stage in the research of FMSs, because they are compatible with III-V based QWs and heterostructures that are used in real semiconductor devices. These (III,Mn)As materials are p-type FMSs, in which Mn atoms act as acceptors and local magnetic moments, and provide unique opportunities to study the hole-induced ferromagnetism, bandstructure, transport, optical properties, and their interplays in well controlled environments. Particularly, (GaMn)As is a prototypical FMS, and lots of experimental and theoretical studies have been concentrated on this material since the late 1990's[22]. However, fundamental understanding and control of the properties of these FMSs for applications remain to be achieved. In particular, there has been a controversy over the bandstructure and the Fermi level position in (GaMn)As which are both closely related to the origin of ferromagnetism[23]. Another remaining problem is that there have been only p-type III-V based FMSs available, because it has been difficult to realize reliable n-type FMS.

In this paper, we review the recent progress in the studies of III-V based FMSs with focus on some important aspects; tunneling transport in (GaMn)As based heterostructure MTJs, and resonant tunneling spectroscopy that leads to the characterization of the band bandstructure and the Fermi level position in (GaMn)As, as described in chapter II. Furthermore, in chapter III, we show the growth of n-type FMS, (InFe)As, and its structural, magnetic, transport, and magneto-optical properties together with electron-induced ferromagnetism.

## II. P-TYPE (GaMn)As AND HETERO-STRUCTURES: TUNNELING TRANSPORT, BANDSTRUCTURE, AND FERMI LEVEL

### A. Tunneling Magnetoresistance (TMR) in GaMnAs-based magnetic tunnel junctions (MTJs) with a single barrier

#### 1. Observation of TMR in GaMnAs-based MTJs

Over the past 15 years, tunneling magnetoresistance (TMR) in magnetic tunnel junctions (MTJs) composed of ferromagnetic semiconductor (FMS) (Ga,Mn)As layers and a nonmagnetic semiconductor tunnel barrier has been extensively studied. (Ga,Mn)As is a III-V based ferromagnetic p-type semiconductor with a zinc-blende-type single crystal having almost the same lattice constant as GaAs and AlAs, and thus heterostructures can be epitaxially grown with abrupt interfaces and with atomically controlled layer thicknesses. There are a lot of advantages in GaMnAs-heterostructures. (i) One can form high-quality single-crystalline MTJs made of all-semiconductor heterostructures, which can be easily integrated with other III-V based structures and devices. (ii) Many parameters such as the barrier height, barrier thickness, and the Fermi energy of ferromagnetic electrodes are controllable. (iii) Introduction of quantum heterostructures, such as double-barrier resonant tunneling diodes (RTDs), is easier than any other material systems.

The first observation of TMR in the GaMnAs-based heterostructure was reported in 1999 by Hayashi et al.[24] in $Ga_{0.961}Mn_{0.039}As$ (200 nm) / AlAs (3 nm) / $Ga_{0.961}Mn_{0.039}As$ (200 nm), showing the TMR ratio of 5% at 4.2 K. Here, we define the TMR ratio as $(R_{AP}-R_{P})/R_{P}$, where $R_{AP}$ and $R_{P}$ are resistances in anti-parallel magnetization and that in parallel magnetization at zero-magnetic field, respectively. Because the lattice constant of GaMnAs is slightly larger than that of GaAs, the GaMnAs film grown on GaAs receives a compressive strain. In this case, the GaMnAs film has an in-plane magnetic anisotropy. Thus, hysteretic MR curves can be obtained when a magnetic field is applied along the in-plane direction of the GaMnAs film. After this report, in 2000, Chiba et al.[25] observed TMR in a $Ga_{0.95}Mn_{0.05}As$ / AlAs (3 nm) / $Ga_{0.97}Mn_{0.03}As$ MTJ grown on an InGaAs buffer layer, whose lattice constant is larger than GaMnAs. In this case, the GaMnAs films receive a tensile strain, leading to an out-of-plane magnetic anisotropy. They applied magnetic field perpendicular to the film plane, and observed the TMR ratio around 5.5% at 20 K.

In 2001, a much higher TMR ratio (>70%, maximum 75%) was observed by Tanaka and Higo in GaMnAs/AlAs/GaMnAs MTJs at 8 K.[26,27] This remarkable increase in TMR was achieved by the improved quality of the GaMnAs layers and the tunnel barrier, and by the optimization of the device structure. It is known that there are many point defects in GaMnAs such as the As antisites[28] and the Mn interstitials[29] whose concentration strongly depends on the molecular beam epitaxy (MBE) growth conditions. Thus, controlling these defects is important for fabricating GaMnAs-based MTJ structures of high quality and high TMR ratios. Later, GaMnAs MTJ structures with a various barrier material such as GaAs,[30] ZnSe,[31] InGaAs,[32] AlMnAs[33] were investigated. It has been known that TMR increases with decreasing temperature, and currently the highest TMR ever reported is 290% at 0.39 K.[30] In addition, the current induced magnetization reversal was demonstrated with very low switching current densities of ~10^5 Acm^{-2}.[32,34]

## 2. Basic properties of TMR in GaMnAs MTJs

Here, we show the basic properties of TMR in the GaMnAs MTJs investigated in our studies, in which we fabricated mesa diode structures schematically shown in Fig. 1 (a) by using conventional photolithography. Figure 1 (b) shows the typical TMR curves of the GaMnAs MTJs, which were observed at 8 K in the mesa diodes with a diameter $\phi$ of 200 μm comprising a Ga$_{1-x}$Mn$_x$As ($x$=4.0%, 50 nm) / AlAs (1.6 nm) / Ga$_{1-x}$Mn$_x$As ($x$=3.3%, 50 nm) trilayer when a magnetic field was applied along the [100] axis in the plane which is one of the easy axes of the GaMnAs layers.[26] Bold solid and dashed curves (major loops) in Fig. 1 (b) were obtained by sweeping the field from positive to negative and negative to positive, respectively. As shown by the bold solid curve, when the magnetic field $H$ was swept from the positive saturation field down to negative, the tunnel resistance $R(H)$ increased from 0.014 to 0.025 Ωcm$^2$ (corresponding TMR was 72 %) at $H$= -110 Oe where the magnetization of one GaMnAs layer reversed and the magnetization configuration changed from parallel to antiparallel. Sweeping the field further to the negative direction, the $R(H)$ and the TMR decreased to the initial value [$R(H)$=0.014 Ωcm$^2$] at $H$= -125 Oe where the magnetization of the other GaMnAs layer reversed and the magnetization configuration became parallel again. A thin solid curve in Fig. 1 (b) shows a minor loop, indicating that the antiparallel magnetization configuration is stable, as well as the parallel magnetization configuration. The switching fields of the major and minor loops are different. This is a general feature of the coercivity in GaMnAs films, and it depends on the maximum magnetic field applied just before the switching. This is probably related to the domain wall pinning, whose unpinning energy depends on the maximum magnetic field applied just before the switching of magnetization. In this case, the minor loop was measured with the maximum magnetic field of -120 Oe, while the major loops were measured with the maximum magnetic fields of ±1 T. Therefore, the switching field of the minor loop was smaller than that of the major loop.

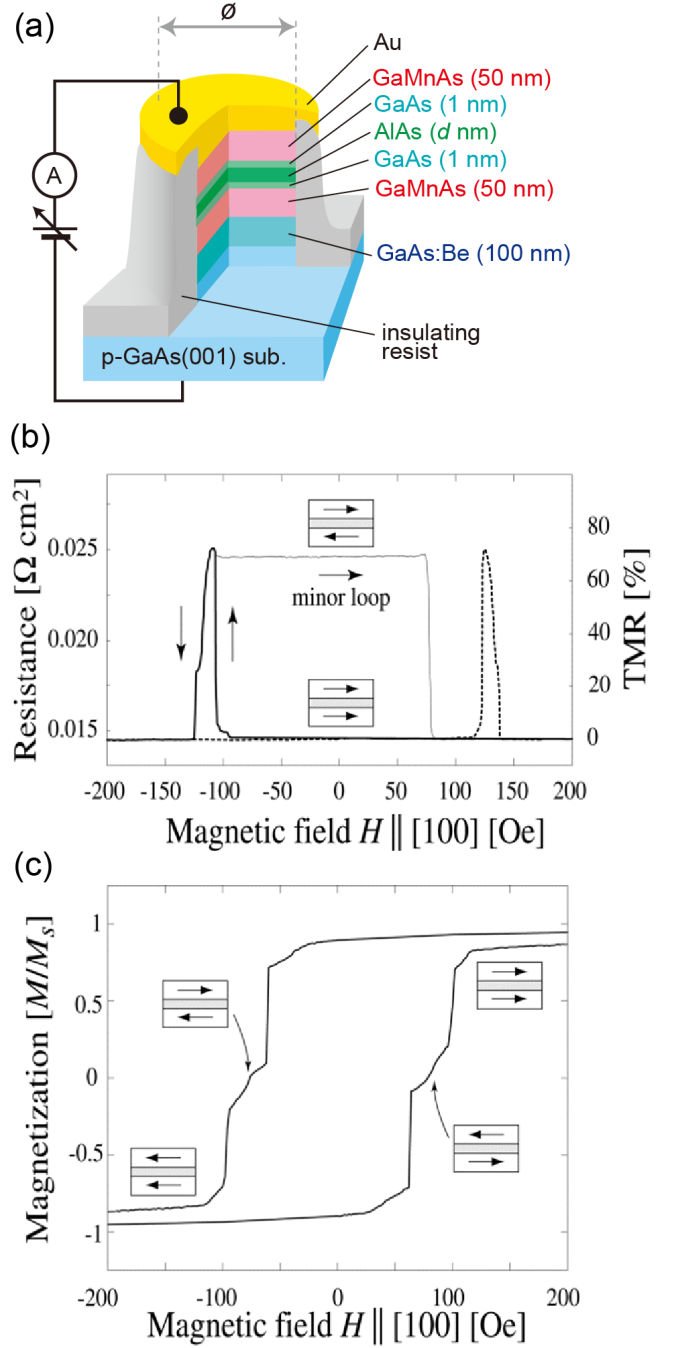

FIG. 1. (Color online) **(a)** Schematic device structure of the GaMnAs MTJ comprising Ga$_{1-x}$Mn$_x$As ($x$=4.0%, 50 nm)/ GaAs (1 nm)/ AlAs ($d_b$ nm)/ GaAs (1 nm)/ Ga$_{1-x}$Mn$_x$As($x$=3.3%, 50 nm)/ GaAs:Be grown on a p$^+$GaAs(001) substrate, where the diameter $\phi$=200 μm. **(b)** TMR curves of this MTJ at 8 K. Bold solid and dashed curves are major loops, with the magnetic field sweep direction from positive to negative and negative to positive, respectively. A minor loop is shown by a thin solid curve. **(c)** Magnetization of this sample with the specimen size of 3 x 3 mm$^2$ at 8 K. The vertical axis shows the normalized magnetization $M/M_s$, where $M_s$ is the saturation magnetization. In both (b) and (c), the magnetic field was applied along the [100] axis in the plane. (b) and (c) are reprinted with permission from Phys. Rev. Lett. **87**, 026602 (2001). Copyright 2001 American Physical Society.

Figure 1 (c) shows the magnetization as a function of magnetic field of this sample measured by superconducting quantum interference device (SQUID) at 8 K. In the SQUID measurements, the sample was cleaved into a square shape with an area of 3 x 3 mm². The magnetic field was applied along the [100] axis in the film plane. Pairs of arrows in the figure indicate the magnetization directions of the top and bottom GaMnAs layers at different fields. Due to the different coercivity of the two GaMnAs layers, a double-step magnetization curve with coercive fields of about 60 Oe and 100 Oe was observed. The difference of the coercive fields between the TMR curves in Fig. 1 (b) and the $M$-$H$ curve in Fig. 1 (c) is caused by the difference in the shape and size of the measured specimens.

The tunnel resistance in the GaMnAs-based single-barrier MTJs can be understood by a conventional tunneling model. Figure 2 shows the resistance area product ($RA$) - $d_b$ characteristics in the GaMnAs MTJs with the AlAs (rectangles),[26] paramagnetic $Al_{0.95}Mn_{0.05}As$ (inverted triangles),[33] and GaAs (circles)[33] barriers, where $d_b$ is the barrier thickness. We found that the paramagnetic AlMnAs is an excellent barrier material; the Curie temperature ($T_C$) of the lower GaMnAs layer can be increased to 60–70 K even when its thickness is thin (2 nm) by using an AlMnAs barrier in the GaMnAs single-barrier MTJ structures (otherwise it is less than ~30 K), probably because the interstitial Mn atoms can more easily pass through the AlMnAs barrier than through the Mn-free barrier during the growth. In all the cases in Fig. 2, the $RA$ exponentially increases with an increase in $d_b$, which means that high-quality tunnel junctions without pinhole current are formed. In the Wentzel–Kramers–Brillouin (WKB) approximation, the slope of $\ln RA$ - $d_b$ characteristics is given by $2[2m^*V_b]^{1/2}/\hbar$, where $m^*$ is the effective mass of holes in the tunnel barrier and $V_b$ the barrier height. If we assume that holes tunnel through the heavy-hole (HH) decaying state in the barrier, $V_b$ is estimated to be 0.45 eV, 0.11eV, and 0.08 eV for the AlAs ($m^*$=0.7$m_0$), AlMnAs ($m^*$=0.7$m_0$), and GaAs ($m^*$=0.45$m_0$) barriers, respectively. Here, $m_0$ is the free-electron mass.

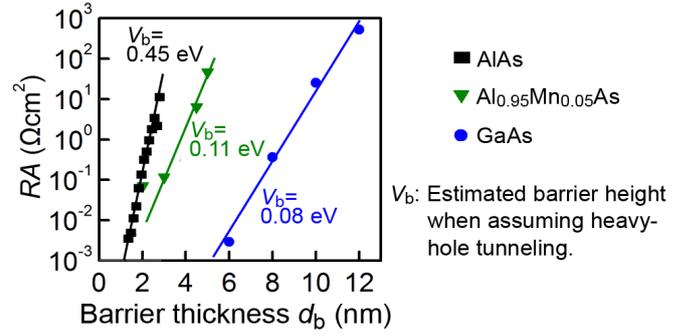

FIG. 2. (Color online) Resistance - area product ($RA$) as a function of the barrier thickness $d_b$ in the GaMnAs-based MTJs when the tunnel barrier is AlAs, $Al_{1-y}Mn_yAs$ ($y$=5%), and GaAs. These data were obtained at zero magnetic field in parallel magnetization with a bias voltage of 1 mV. Reprinted with permission from Appl. Phys. Lett. **95**, 242503 (2009). Copyright 2009 American Institute of Physics.

Figure 3 (a) shows the $d_b$ dependence of TMR in the GaMnAs MTJs with the tunnel barrier of AlAs (black circles) and paramagnetic $Al_{1-y}Mn_yAs$ (blue diamonds for $y$=5% and red triangles for $y$=12%) when a magnetic field is applied along the in-plane [100] axis. In all the cases, the TMR values are suppressed in the small

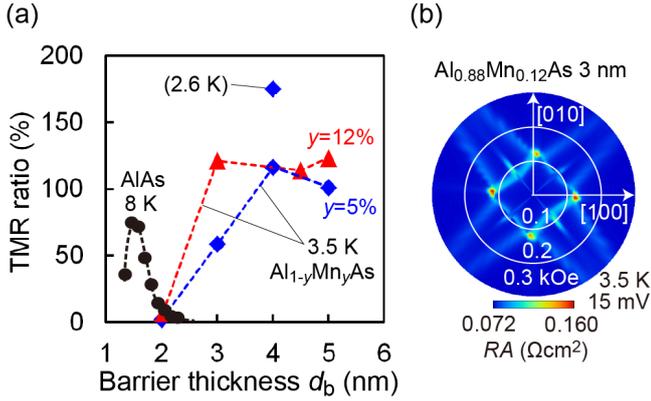

**(a)**

**(b)**

**FIG. 3.** (Color) **(a)** Barrier thickness $d_b$ dependence of TMR in the GaMnAs MTJs with an AlAs, $Al_{1-y}Mn_yAs$, and GaAs barrier, where the bias voltage was set so that the TMR became maximum in each device. [As a reference, a typical bias dependence of TMR is shown in Fig. 4(b).] The measurement temperature was 8 K, 3.5 K, and 3.5 K, respectively. Reprinted with permission from Appl. Phys. Lett. **95**, 242503 (2009). Copyright 2009 American Institute of Physics. **(b)** Magnetic-field dependence of $RA$ in $Ga_{0.94}Mn_{0.06}As/ Al_{0.88}Mn_{0.12}As$ (3 nm)/ $Ga_{0.94}Mn_{0.06}As$ MTJ as a function of the in-plane magnetic-field direction.

$d_b$ region ($d_b < 1.5$ nm for AlAs, $d_b < 3$ nm for AlMnAs), which is considered to be due to the magnetic coupling between the two GaMnAs layers. With increasing $d_b$, the TMR value decreases in the case of the AlAs barrier, whereas it is almost maintained in the case of the AlMnAs barrier. The different features between AlAs and AlMnAs are probably due to the difference of the crystallinity between them. In the MBE growth of these heterostructures, low growth temperature at ~200°C is necessary in order to prevent the phase separation of GaMnAs. During the MBE growth, the reflection high energy electron diffraction (RHEED) of AlMnAs was a more streaky 1×1 pattern than that of AlAs when the substrate temperature was below 200°C, which led to the higher TMR ratios when the AlMnAs barrier was used. If we can improve the crystal quality of the low-temperature grown AlAs tunnel barrier, a very large TMR will be obtained. Another possibility is that the TMR is enhanced by magnetic-impurity assisted tunneling.[35] In this case, the interstitial and substitutional Mn in the AlMnAs barrier can enhance the TMR, though this effect may be small because the energy of the Mn impurity states are far from the Fermi level in almost all the AlMnAs barrier region.[33]

TMR in GaMnAs MTJs has a characteristic feature of the in-plane magnetic-field angle dependence, which is induced by an in-plane magnetic anisotropy of GaMnAs mainly expressed by the sum of the two components; biaxial easy axes (along the [100] and [010] directions) and a uniaxial easy axis (along [110] or [1$\bar{1}$0]).[36] Figure 3(b) shows a typical magnetic-field dependence of $RA$ (TMR curve) in a GaMnAs MTJ with a paramagnetic AlMnAs tunneling barrier as a function of the magnetic-field direction in the plane. Here, before starting to measure TMR, a strong magnetic field of 1 T was applied in the direction opposite to the intended magnetic-field direction and then was reduced to zero. After that, the measurement was started with increasing magnetic field in the intended direction. The highest TMR values were obtained in the four regions at ~0.13 kOe near the four easy magnetization axes ([100], [010], [$\bar{1}$00], and [0$\bar{1}$0]) of the GaMnAs layers, which means that nearly perfect anti-parallel magnetization alignment is achieved there. We note that this feature is completely different from tunneling *anisotropic* magneto resistance (TAMR), in which the sign of the TAMR ratio is reversed when the magnetic field direction is rotated by 90° in the plane,[37] whereas the TMR ratio never becomes negative in any of the magnetic-field directions as shown in the $RA$ mapping in Fig. 3 (b).

Figure 4 (a) shows the example of the temperature dependence of TMR of the GaMnAs-based MTJ with a 4-nm-thick $Al_{0.95}Mn_{0.05}As$ tunnel barrier, measured with a magnetic field applied along the in-plane [100] axis when a bias voltage is 10 mV. We see that the TMR signal persists up to 60 K, which is consistent with $T_C$ of the GaMnAs layers in this MTJ. TMR increases rapidly with decreasing temperature, which is a typical feature of the GaMnAs MTJs, although the clear reason has not been clarified yet. Figure 4 (b) shows the examples of the bias dependences of TMR, which were obtained in the GaMnAs MTJs with an Al$_{0.88}$Mn$_{0.12}$As barrier whose thickness was 3 nm (red), 4.5 nm (green), and 5 nm (blue). The measurements were carried out at 3.5 K with a magnetic field applied in plane along the [100] axis. Usually, in GaMnAs MTJs, an increase in the bias voltage monotonically suppresses TMR, which is similar to the typical characteristic of the metal-based MTJ devices. In the magnified view shown in the inset of Fig. 4 (b), however, we see that the AlMnAs-barrier devices show a little different feature; TMR is suppressed near zero bias especially in the negative bias region (where holes are injected from the bottom GaAs:Be electrode) in all the data. It is plausible to think that the singular behavior of the TMR is due to resonant tunneling through the impurity states of the Mn atoms in the AlMnAs barrier, which are located about 500 meV higher than the valence band (VB) of top of AlMnAs.[38] Holes are affected by these impurity levels only near the interfaces in the AlMnAs barrier because the energy of these levels is close to the Fermi level due to the band bending of the barrier near the interfaces (see Fig. 1 in Ref. 33). When $d_b$ gets thinner, this effect becomes larger compared with the total tunneling sequence, thus suppressing the TMR near zero bias more strongly.

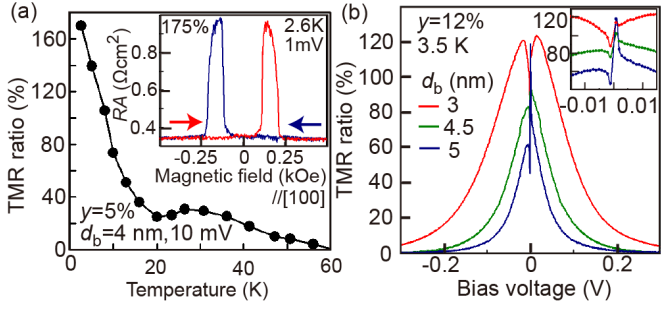

**FIG. 4.** (Color) **(a)** Inset shows the magnetic-field dependence of $RA$ of the MTJ with a 4-nm-thick Al$_{0.95}$Mn$_{0.05}$As barrier at 2.6 K with a magnetic field applied in the plane along the [100] axis when the bias voltage is 1 mV. The main graph shows the temperature dependence of TMR of this device with a magnetic field applied in the plane along the [100] axis when the bias voltage is 10 mV. **(b)** Bias dependence of TMR of the GaMnAs-based MTJs with an Al$_{0.88}$Mn$_{0.12}$As barrier thickness of 3 nm (red), 4.5 nm (green), and 5 nm (blue) when a magnetic field is applied in the film plane along the [100] axis. The inset shows the magnified view near zero bias. Reprinted with permission from Appl. Phys. Lett. **95**, 242503 (2009). Copyright 2009 American Institute of Physics.

## B. Spin-dependent resonant tunneling in GaMnAs-based quantum heterostructures

### 1. Observation of resonant tunneling and TMR increase in double-barrier MTJs with a GaMnAs quantum well[39]

The characteristic features of GaMnAs heterostructures such as high-quality single crystallinity, flat interfaces, and good compatibility to III-V heterostructures, are quite attractive and have motivated us to investigate quantum heterostructures with spin-degrees of freedom in order to achieve new functionalities. Combining the quantum-size effect and TMR is one of the very important steps for this purpose. A very large enhancement of TMR up to 800 % and to more than $10^6$ % has been theoretically expected in GaMnAs-based RTD structures,[40,41] in which the spin-split resonant levels in the ferromagnetic quantum well (QW) are used as an energy filter as well as a sharp spin-filter. In magneto-optical measurements of the GaMnAs QW, blue shifts of the magneto-optical spectra were observed, suggesting the existence of the quantum-size effect in GaMnAs.[42,43]

Clear observation of spin-dependent resonant tunneling in the GaMnAs QW was first demonstrated in 2007.[39] The device structure was Ga$_{0.95}$Mn$_{0.05}$As(20 nm)/ GaAs(1 nm)/ Al$_{0.5}$Ga$_{0.5}$As(4 nm)/ GaAs(1 nm)/ Ga$_{0.95}$Mn$_{0.05}$As($d$ nm)/ GaAs(1 nm)/ AlAs(4 nm)/ Be-doped GaAs(100 nm) grown by MBE on a p$^+$GaAs(001) substrate, where the thickness $d$ of the GaMnAs

QW ranges from 3.8 to 20 nm. The Be concentration of the Be-doped p-type GaAs (GaAs:Be) layer was $1 \times 10^{18}$ cm$^{-3}$. The 1-nm-thick GaAs layers were inserted to prevent the Mn diffusion into the barrier layers and to smooth the interfaces. The GaAs:Be, AlAs, and the lowest GaAs spacer layers were grown at 600, 550, and 600 ℃, respectively. The GaMnAs layers were grown at 225 ℃, and GaAs/ AlGaAs/ GaAs layers below the top GaMnAs layer were grown at 205 ℃. We grew four samples named A, B, C, and D. When growing sample A, we moved the in-plane position of the main shutter equipped in our MBE chamber in front of the sample surface during the growth of the GaMnAs QW, and varied $d$ from 3.8 to 8.0 nm on the same sample wafer. In the growth of samples B, C, and D, $d$ was fixed at 12, 16, and 20 nm, respectively. The device fabrication process was the same as that mentioned in the previous section with the mesa diameter $\phi$ of 200 μm. In the following measurements, the bias polarity is defined by the voltage of the top GaMnAs electrode with respect to the substrate.

The schematic band diagrams when negative and positive biases $V$ are applied to the heterostructure are shown in Fig. 5 (a) and (b), respectively. In this structure, TMR occurs by tunneling of holes between the ferromagnetic GaMnAs top electrode and the ferromagnetic GaMnAs QW. The following tunneling transport measurements were carried out in a cryostat cooled at 2.6 K with a conventional two-terminal direct-current (DC) method. $dI/dV$-$V$ and $d^2I/dV^2$-$V$ characteristics were derived mathematically from the data of the $I$-$V$ characteristics measured at every 5 mV. The results of the bias dependence of TMR were obtained from the data of $I$-$V$ characteristics measured in parallel and anti-parallel magnetizations. $I$-$V$ characteristics in both parallel and anti-parallel magnetizations were obtained at zero-magnetic field.

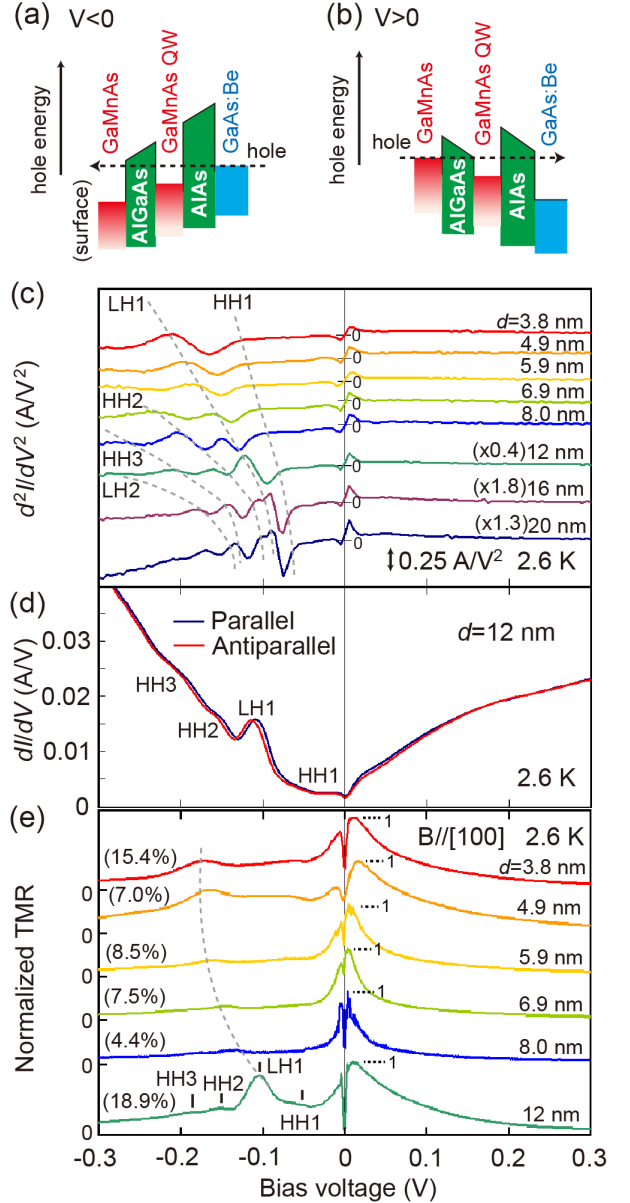

**FIG. 5.** (Color) **(a) (b)** Schematic band diagrams of the RTD junction when the bias polarity is negative and positive, respectively. Here, the 1-nm-thick GaAs spacer layers are omitted for simplicity. **(c)** $d^2I/dV^2$-$V$ characteristics of these RTD junctions with various QW thicknesses $d$ in parallel magnetization at 2.6 K. Numbers in the parentheses express the magnification ratio for the vertical axis. **(d)** $dI/dV$-$V$ characteristics of the junction with $d = 12$ nm at 2.6 K in parallel (blue curve) and antiparallel (red curve) magnetization. **(e)** Bias dependence of TMR in Ga$_{0.95}$Mn$_{0.05}$As(20 nm)/ GaAs(1 nm)/ Al$_{0.5}$Ga$_{0.5}$As(4 nm)/ GaAs(1 nm)/ Ga$_{0.95}$Mn$_{0.05}$As($d$ nm)/ GaAs(1 nm)/ AlAs(4 nm)/ GaAs:Be(100 nm) RTD junctions with various QW thicknesses $d$ when a magnetic field was applied in the plane along the [100] direction at 2.6 K, where the TMR ratios are normalized by the maximum value of TMR in each curve shown in the parenthesis. Reprinted with permission from Phys. Rev. B **75**, 155328 (2007). Copyright 2007 American Physical Society.

Figure 5 (c) shows $d^2I/dV^2$-$V$ characteristics of the junctions in parallel magnetization at 2.6 K. Sharp features near the zero bias are observed in all the curves with various $d$, corresponding to the zero-bias anomaly which is usually observed in GaMnAs-based heterostructures.[44] The most important feature in Fig. 5 (c) is the oscillations whose peak voltages depend on $d$ in the negative bias region of all the curves. As $d$ increases, these peaks shift to smaller voltages and the period of the oscillation becomes short. Such oscillatory behavior has not been observed in GaMnAs-based single-barrier MTJs, indicating that the resonant tunneling effect induces the oscillations. (Later, we explain more clearly that no other mechanisms can explain the oscillations observed in our studies.) As shown in Fig. 5 (c), by the theoretical analysis mentioned below, the peaks HH$n$ and LH$n$ ($n$=1, 2, 3 ···) are assigned to resonant tunneling through the $n$th level of the heavy hole (HH) band and light hole (LH) band in the GaMnAs QW, respectively. Figure 5 (d) shows the $dI/dV$-$V$ curves of the junction with $d$=12 nm at 2.6 K in parallel (blue curve) and anti-parallel (red curve) magnetization. These two curves are almost the same, but there is a little voltage shift (7 mV) which is most obvious at LH1.

TMR was clearly observed in the junctions with $d$ from 3.8 to 12 nm. Figure 5 (e) shows the bias dependence of the TMR ratio of these junctions at 2.6 K with a magnetic field applied in plane along the [100] direction, where the TMR ratios are normalized by the maximum value of TMR in each curve shown in the parenthesis. TMR oscillations can be seen in the negative bias region of all the curves. With increasing $d$, the TMR peaks (except for that near zero bias) shift to smaller voltages as is the case of the $d^2I/dV^2$-$V$ characteristics shown in Fig. 5 (c), which indicates that the TMR increase is induced by resonant tunneling. A significant TMR increase occurs at LH1 (especially when $d$ =12 nm), which is caused by the magnetization-dependent peak's shift at LH1 observed in the $dI/dV$-$V$ characteristics shown in Fig. 5 (d).

To understand the resonant peaks observed in this study, we calculated the quantum levels of GaAs/ Al$_{0.5}$Ga$_{0.5}$As(4 nm)/ GaMnAs($d$ nm)/ AlAs(4 nm)/ GaAs by using the transfer matrix method[45] with the 4×4 $k \cdot p$ Hamiltonian[46] and the $p$-$d$ exchange Hamiltonian [47] for introducing the in-plane magnetization of the GaMnAs QW in the assumed VB lineup shown in Fig. 6 (a). The splitting energy $\Delta$ of the GaMnAs QW, which corresponds to the spin splitting energy for the LH at the $\Gamma$ point, is set at (3 meV, 0, 0) along the in-plane [100] direction parallel to the magnetic field applied in our experiments. The center energy of the spin-split VB of the GaMnAs QW was located at 28 meV above the VB of GaAs in terms of hole energy. We note that this band lineup *contradicts* the conventional VB conduction picture of GaMnAs, where it is assumed that the Fermi level exists in VB. In Fig. 5 (c), the resonant peaks tend to converge at ~ -60 mV, which corresponds to the VB top position because the resonant levels are formed by the quantization of VB. As can be seen in Fig. 6 (a), this fact means that the Fermi level of GaMnAs exits in the band gap. (We will explain it in detail in the later sections.) Figure 6 (b) shows the calculated result of the resonant-peak bias voltages for the HH resonant levels (red points and curves assigned to HH$n$) and the LH resonant levels (blue points and curves assigned to LH$n$) at $k_\parallel$=0, where $k_\parallel$ is the wave vector parallel to the film plane. These resonant peak bias voltages were derived by multiplying the calculated energy values of the quantum levels by $s$=2.5 (ideally, $s$ is 2.[48]) for better fit, which means that 20% [=(2.5-2)/2.5] of the applied bias voltage is consumed in the electrodes. Since the HH spins are oriented along the tunneling direction and the $p$-$d$ exchange Hamiltonian is proportional to $s \cdot S$, the quantum levels of HH are not spin split but those of LH are split by the in-plane magnetization,[49] where *s* and *S* are spins of the carrier and the Mn atom, respectively.  Holes tunnel through the lower and upper states of the spin-split LH quantum levels in parallel and in antiparallel magnetization, respectively.  The calculated voltages of the LH quantum levels shown in Fig. 6 (b) are those in parallel magnetization.

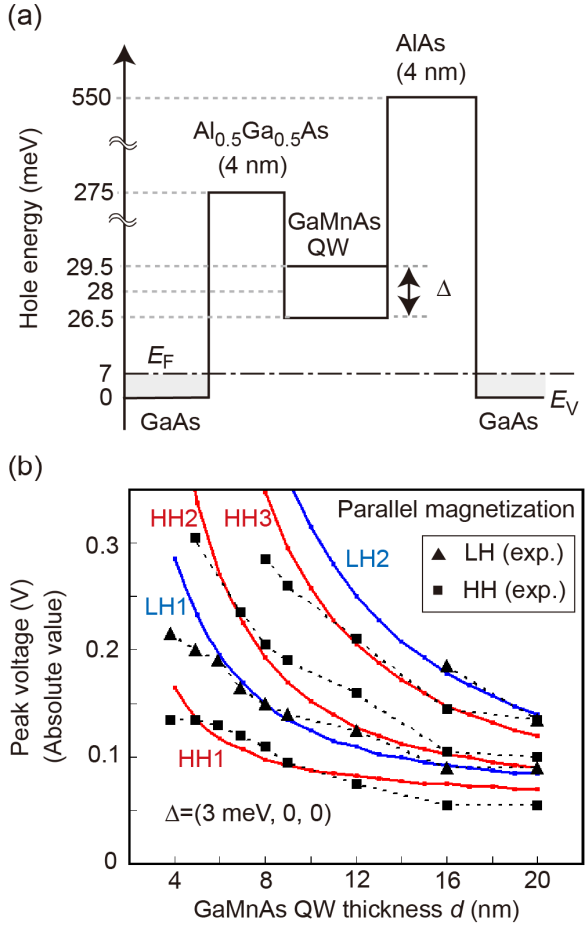

(a)

(b)

**FIG. 6.** (Color online)  **(a)** Valence-band diagram assumed in our calculation of the quantum levels in GaAs/ Al$_{0.5}$Ga$_{0.5}$As(4 nm)/ GaMnAs($d$ nm)/ AlAs(4 nm)/ GaAs. $E_F$, $E_V$, and $\Delta$ mean the Fermi level, the valence band edge, and the spin-splitting energy of the light-hole band of the GaMnAs QW at the $\Gamma$ point, respectively.  **(b)** Calculated and experimentally obtained resonant peak voltage *vs.* the GaMnAs QW thickness $d$.  The black solid rectangles and triangles denote the experimentally obtained resonant-peak voltages assigned as HH and LH quantum levels in the $d^2I/dV^2$-$V$ characteristics of Ga$_{0.95}$Mn$_{0.05}$As(20 nm)/ GaAs(1 nm)/ Al$_{0.5}$Ga$_{0.5}$As(4 nm)/ GaAs(1 nm)/ Ga$_{0.95}$Mn$_{0.05}$As($d$ nm)/ GaAs(1 nm)/ AlAs(4 nm)/ GaAs:Be(100 nm) RTD junctions in parallel magnetization, respectively.  Here, these voltages are expressed in the absolute values.  The small red and blue points (curves) denote the calculated resonant voltages of HH and LH, respectively.  The quantum levels of LH1 and LH2 are spin-split by the in-plane magnetization introduced by the *p-d* exchange Hamiltonian.  The calculated voltages of the LH quantum levels shown in this figure are those in parallel magnetization.  Reprinted with permission from Phys. Rev. B **75**, 155328 (2007). Copyright 2007 American Physical Society.

Figure 6 (b) also shows the experimental peak voltages assigned to HH and LH quantum levels observed in the $d^2I/dV^2$-$V$ curves in parallel magnetization by black solid rectangles and triangles, respectively.  Here, the peak voltages are expressed in the absolute values.  Although there is a little deviation between the experimental and calculated results, the experimental results are well fitted by the present model.  Possible origins of the small deviations are considered as follows.  When *d* is thinner than 5 nm, Mn diffusion from the GaMnAs QW layer to the adjacent GaAs spacer layer yields lowering of the resonant tunneling energy.  The deviations may also come from the point defects incorporated into GaMnAs such as Mn interstitials[29] and As anti-site defects,[28] whose concentrations are very sensitive to the growth condition of GaMnAs.[50]  These defects can influence the spin-splitting energy, band offset, and strain, leading to such deviations.

In our experiment, the resonant peaks were observed only in the negative bias region, where holes are injected from the p-type GaAs:Be layer to the GaMnAs QW.  This is attributed to the relatively weak resonant tunneling (energetically broad resonant levels) and the difference of the Fermi surface areas between the top GaMnAs and the bottom GaAs:Be electrodes.[39]  Figures 7 (a) and (b) show the hole subband structures of AlAs(1nm)/ GaAs QW (5nm)/ AlAs (1 nm) calculated by the *k·p* model.  For simplicity, we neglected the spin splitting.  The GaMnAs electrode has a high hole concentration ($\sim 10^{21}$ cm$^{-3}$), that is, a large Fermi surface, meaning that the holes tunnel in a large $k_\parallel$ region when they are injected from GaMnAs.

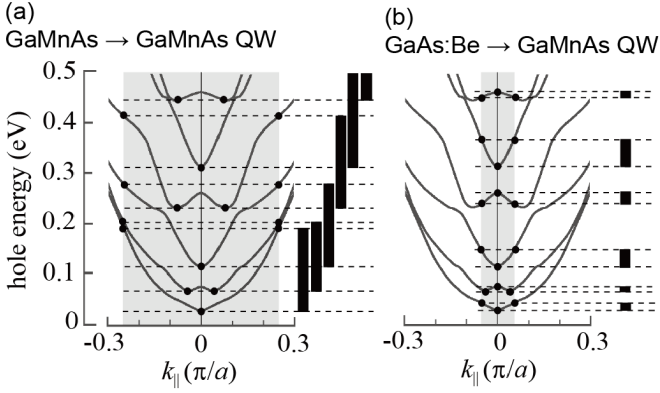

(a)
GaMnAs → GaMnAs QW

(b)
GaAs:Be → GaMnAs QW

**FIG. 7.** **(a)** **(b)** Schematic hole-subband structures of AlAs(1nm)/ GaAs QW (5nm)/ AlAs (1 nm) calculated by the **k·p** model. The gray regions correspond to the $k_{\parallel}$ regions where holes are injected from the (a) GaMnAs and (b) GaAs:Be, and the black bands on the right side are corresponding energy regions. The circular points correspond to the states of the maximum and minimum energies in each subband within the gray region. The broken lines are maximum and minimum energies of each subband within the gray region. Reprinted with permission from Phys. Rev. B **75**, 155328 (2007). Copyright 2007 American Physical Society.

In contrast, holes tunnel in a small $k_{\parallel}$ region when they are injected from GaAs:Be with a lower hole concentration ($\sim 10^{18}$ cm$^{-3}$) and a smaller Fermi surface. Here, we assume that holes tunnel in the gray regions of VB shown in Figs. 7 (a) and (b) when they are injected from GaMnAs and GaAs:Be, respectively. The black circular points correspond to the states of the maximum or minimum energies in each subband in the gray region. The black bands shown on the right side are the energy regions corresponding to these subbands in the gray region. We see that these energy regions of the subbands are energetically overlapped in (a). Furthermore, these subbands are broadened by a lot of scattering in GaMnAs, so it is very difficult to detect these subbands separately in tunneling transport measurements when carriers are injected from GaMnAs. Meanwhile, when carriers are injected from GaAs:Be, the energy regions of the subbands are energetically separated. Thus, the resonant peaks are observed only in the negative bias region. Our result indicates that electrodes with a low carrier concentration are appropriate for clear detection of the resonant tunneling effect in GaMnAs QW heterostructures.

## 2. Enhanced resonant tunneling and TMR in RTDs with a thin GaMnAs QW[51]

Although we successfully observed the resonant tunneling effect and TMR increase in GaMnAs QW, the huge TMR enhancement predicted in the theories ($\sim$800%,[40] $\sim 10^6$%[41]) was not obtained.[39] This is because a large VB spin splitting ($\sim$100 meV) was assumed in these theories, while in fact we observed only a small spin splitting (3 meV) of the LH1 level. We think that a part of the reason was the poor quality of the GaMnAs QW, where its $T_C$ was only 30 K due to the difficulty in removing the Mn interstitial defects in the GaMnAs QW because it is sandwiched by the barriers. (Later, we found that the low $T_C$ was *not* an essential cause of this problem of the small spin splitting, as will be described in the next section.) We found that the following two ways are effective to increase $T_C$ and to enhance resonant tunneling to a certain degree. One is to reduce the QW thickness for inducing stronger quantization, and the other is to use paramagnetic AlMnAs as an upper tunnel barrier of the GaMnAs QW. As described above, we found that $T_C$ of the lower GaMnAs layer can be increased to 60-70 K even when its thickness is thin (2 nm) by using a paramagnetic AlMnAs tunnel barrier in the GaMnAs single-barrier MTJ structures.[33] This improvement is probably because the interstitial Mn atoms can more easily pass through the AlMnAs barrier than through the Mn-free barrier during the growth.

As shown in Fig. 8 (a), we grew Ga$_{0.94}$Mn$_{0.06}$As(10 nm)/ GaAs(1 nm)/ (Al$_{0.5}$Ga$_{0.5}$As or Al$_{0.94}$Mn$_{0.06}$As) (4 nm) / GaAs(2 nm)/ Ga$_{0.94}$Mn$_{0.06}$As(2.5 nm)/ GaAs(1 nm)/ AlAs(4 nm)/ Be-doped GaAs(100 nm) on a p$^+$GaAs(001) substrate. We prepared two RTD samples named as RTD A and B with the upper tunnel barrier composed of $Al_{0.5}Ga_{0.5}As$ and $Al_{0.94}Mn_{0.06}As$, respectively. The Be concentration of the Be-doped p-type GaAs (GaAs:Be) layer was $2×10^{18}$ $cm^{-3}$. Figures 8 (b) and (c) show the $dI/dV$-$V$ and $d^2I/dV^2$-$V$ characteristics of the RTD devices fabricated in this study, respectively. In each graph, blue and red (green and orange) curves are the data of RTD A (B). The blue or green curve corresponds to the parallel magnetization, while the red or orange curve corresponds to the antiparallel magnetization. In Fig. 8 (b), oscillations induced by resonant tunneling are observed in the negative bias region in both RTD devices. The resonant peaks are more clearly seen in Fig. 8 (c). The shapes of the $d^2I/dV^2$-$V$ curves of both RTD A and B are quite similar to those observed in other RTD devices with various GaMnAs QW thicknesses described earlier (Fig. 5 (c)),[39] and thus these resonant peaks can be assigned as described in Figs. 8 (b) and (c). The resonant peaks in RTD A are observed at larger voltages than those in RTD B, which is due to the barrier height difference between $Al_{0.5}Ga_{0.5}As$ (~275 meV) and $Al_{0.94}Mn_{0.06}As$ (~110 meV). The peak positions are different between parallel and antiparallel magnetization only at the LH1 state both in RTD A and B, which is consistent with the characteristics of the *p-d* exchange interaction with the in-plane magnetization of GaMnAs.[49] The difference in the resonant peaks' voltage at LH1 between parallel and anti-parallel magnetization is 8 and 13 mV in RTD A and B, respectively. These values are related to the difference of the $T_C$ values of the GaMnAs QW: 30 K (RTD A) and 60 K (RTD B) and correspond to the spin splitting energy of 3 and 5 meV, respectively, which were estimated by the similar quantum-level calculations mentioned above. In Figs. 8 (b) and (c), the resonant level of HH1 is observed both in RTD A and B, which indicates that HH1 is not occupied by holes and the Fermi level of GaMnAs lies in the band gap. This again contradicts the conventional understanding of the VB conduction picture of GaMnAs.

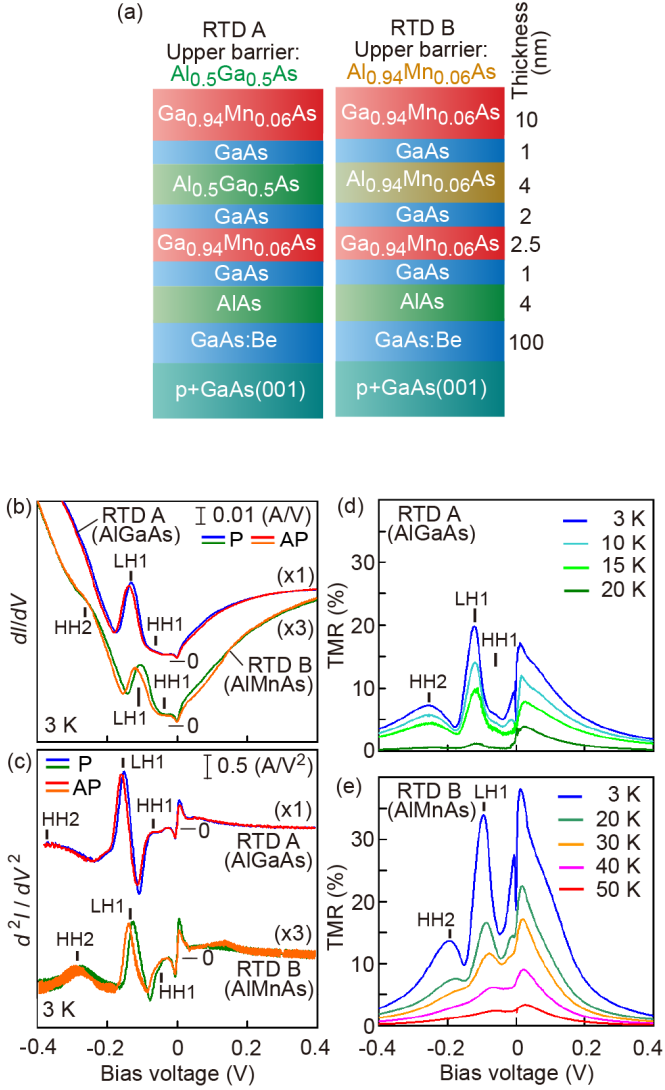

FIG. 8. (Color) **(a)** Schematic cross-sectional structures of RTD A and B. **(b)** $dI/dV$-$V$ and **(c)** $d^2I/dV^2$-$V$ characteristics of RTD A and B at 3 K. In both graphs, the blue and red (green and orange) curves correspond to RTD A and B, respectively. The blue or green curve corresponds to the parallel magnetization (P). The red or orange curve corresponds to the anti-parallel magnetization (AP). **(d)** and **(e)** show the temperature evolution of the bias dependence of TMR obtained in RTD A and B, respectively. Reprinted with permission from Phys. Rev. Lett. **104**, 167204 (2010). Copyright 2010 American Physical Society.

Figures 8(d) and (e) show the temperature evolution of the bias dependence of TMR obtained in RTD A and B, respectively. In both RTD devices, clear enhancement of TMR is seen especially at LH1 reflecting the spin splitting at LH1 as shown in (b) and (c). It is a quite unique property that the TMR ratio at LH1 is *higher* than that at zero bias in RTD A at 3-15 K. Such clear TMR enhancements induced by resonant tunneling are due to the high crystallinity of GaMnAs and have never been observed in the studies of MTJs of other material systems.[52,53] TMR disappears at 30 K in RTD A and at 60 K in RTD B with increasing temperature, which indicates that the $T_C$ of the GaMnAs QW is estimated to be 30 K and 60 K in RTD A and B, respectively. We note that the top GaMnAs layer has a $T_C$ of 60 K which was determined by the magnetic circular dichroism (MCD) measurements.

We succeeded in enhancing TMR, but the TMR ratios are still lower than the theoretically predicted values. Moreover, the spin-splitting energy of LH (3-5 meV) was much smaller than the assumed ones (~100 meV) in the conventional band picture of GaMnAs.[47] To clarify the reason, we came up with an idea to more systematically investigate the VB structure of various GaMnAs films, as described in the next section.

## 3. Resonant tunneling spectroscopy and the VB structure of (III,Mn)As (III=Ga,In)[54,55]

There has been a dispute on the band structure of GaMnAs. The conventional mean-field Zener model, where the ferromagnetism is induced by the *p-d* exchange interaction between the VB holes whose concentration is $10^{20}$-$10^{21}$ cm$^3$ and the localized Mn-3d electrons, has been generally accepted, because it seems to be able to explain a variety of features of GaMnAs.[47,56,57,58,59] In this picture, it is assumed that the VB is merged with the Mn-induced impurity band (IB). The Fermi level position is determined by the concentration of the VB holes and lies 200-300 meV below the top of the merged VB. In contrast, recent reports on the optical[60,61,62,63] and transport[64,65] properties of GaMnAs have shown that the Fermi level exists in the IB[66] within the band gap of GaMnAs [see Fig. 13 (a)]. In this case, the Zener double-exchange-type mechanism is applicable for GaMnAs, where the ferromagnetism is stabilized by hopping of the spin-polarized holes in the IB.[67] Meanwhile, a recent scanning tunneling microscope analysis has shown the microscopically inhomogeneous nature of the electronic structure of GaMnAs, which complicates the understanding of this material.[68] This controversial situation means that for further development of this material, it is strongly needed to clarify the VB picture of GaMnAs.

Resonant tunneling spectroscopy is a very powerful method for characterizing the band structure of FMSs. The resonant levels contain an abundance of information related to material parameters, such as the Fermi level, effective mass (that is, band dispersion), band offset, and strain. Thus, this method gives us useful clues to the mechanism behind the ferromagnetism as well as to the precise band picture. However, it is very difficult to control the properties of GaMnAs QW sandwiched by barriers, because the interstitial Mn atoms cannot be removed out easily through the barriers, reducing the $T_C$ of the GaMnAs QW significantly. Here, we show that this method can be applied also for the *surface* GaMnAs layers.[54] This new finding allows us to systematically investigate a variety of GaMnAs films with a wide range of Mn content and $T_C$, where the characteristics of the surface GaMnAs can be easily controlled by low-temperature annealing.[69]

To investigate the VB structures of GaMnAs, we have used mesa diodes with the diameter $\phi$ of 200 μm composed of Ga$_{1-x}$Mn$_x$As (*d* nm)/ AlAs (5 nm)/ GaAs:Be (100 nm, Be concentration: $1\times10^{18}$ cm$^{-3}$) grown on p$^+$GaAs(001) substrates, as shown in Fig. 9 (a), where the thickness *d* was precisely controlled by changing the etching time from mesa to mesa. Fig. 9 (b) depicts the idealized behavior of the resonant levels formed in the surface GaMnAs layer as *d* is increased from $d_1$ to $d_2$ ($d_1 < d_2$).

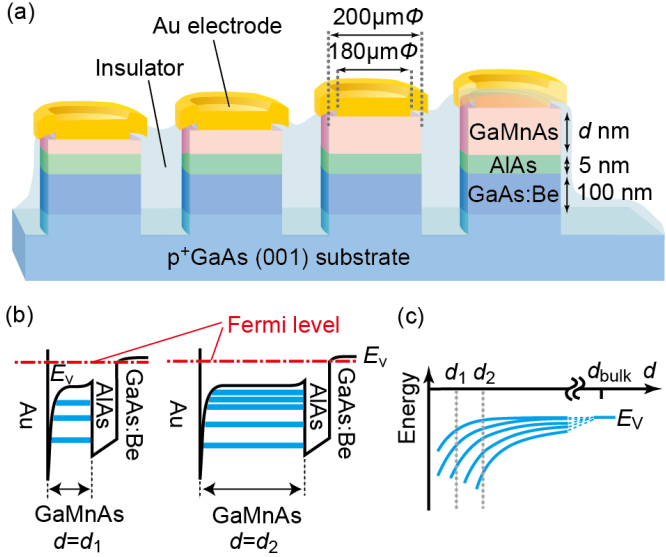

(a) Au electrode
Insulator
200μmΦ
180μmΦ
GaMnAs  $d$ nm
AlAs  5 nm
GaAs:Be  100 nm
p⁺GaAs (001) substrate

(b) Fermi level
$E_V$
Au
GaMnAs
$d = d_1$
AlAs
GaAs:Be
Au
GaMnAs
$d = d_2$
AlAs
GaAs:Be
$E_V$

(c) Energy
$d_1$ $d_2$
$d_{bulk}$ $d$
$E_V$

**FIG. 9.** (Color) **(a)** Schematic device structure composed of GaMnAs ($d$ nm)/AlAs (5 nm)/GaAs:Be (Be:1x10$^{18}$cm$^{-3}$, 100 nm)/p⁺GaAs(001) junctions. **(b)** Schematic band diagrams of the Ga$_{1-x}$Mn$_x$As ($d$ nm)/ AlAs (5 nm)/ GaAs:Be heterostructure when $d = d_1$ and $d = d_2$ ($d_1 < d_2$). The black solid curves, red dash–dotted lines and blue lines indicate the energy of the VB top $E_V$ (neglecting the quantum-size effect), the Fermi level, and the resonant levels of the VB holes, respectively. The strain and exchange splitting are neglected for simplicity. These resonant levels are formed by the VB holes confined by the AlAs barrier and the surface Schottky barrier induced by Fermi level pinning at the surface. **(c)** Ideal $d$ dependence of the resonant level energies. The converged energy corresponds to $E_V$ in the bulk GaMnAs. Reprinted with permission from Nature Phys. **7**, 342 (2011). Copyright 2011 Nature Publishing Group.

The black solid curves, red dash-dotted lines, and blue lines indicate the energy of the VB top $E_V$ (neglecting the quantum-size effect), the Fermi level, and the resonant levels, respectively. The strain and exchange splitting are neglected for simplicity. These resonant levels are formed by the holes confined by the AlAs barrier and the surface Schottky barrier. The surface Schottky barrier is induced by Fermi-level pinning at the surface and its thickness is estimated to be 0.7 - 1.2 nm depending on the hole concentration. These resonant levels can be detected only when holes are injected from the GaAs:Be layer because the injected holes have small in-plane wave vectors $k_∥$ owing to the low hole concentration ($1×10^{18}$ cm$^{-3}$) of the GaAs:Be electrode. In our analyses, energy bands mostly with the symmetry of the p-orbitals are detected in GaMnAs, reflecting the p-character of the holes injected from the GaAs:Be electrode. Ideally, the resonant levels in the surface GaMnAs layer converge at $E_V$ of bulk GaMnAs as $d$ increases [Fig. 9 (c)]. The bias voltage $V$ is defined as the voltage of the top electrode with respect to the substrate.

Figures 10 (a) and (b) show the $d^2I/dV^2$-$V$ curves (black solid curves) obtained for various $d$ measured at 3.4 K in the Ga$_{1-x}$Mn$_x$As ($x$=6%) diodes with $T_C$ of 71 K and 111 K, respectively. In both of the samples, oscillations due to resonant tunneling are clearly detected. Here, the resonant tunneling levels were assigned by the theoretical analyses mentioned below. The HH1 level was observed in both samples, indicating that it is not occupied by holes in the equilibrium condition. As HH1 is the highest VB resonant level (the lowest level in terms of hole's energy) and the Fermi level position corresponds to zero bias, our results clearly indicate that the Fermi level lies in the band gap in both samples. These resonant oscillations persist up to 30-60 K, which contrasts with the weak coherent feature of the holes near the Fermi level reported at low temperatures of less than 1 K.[70] These results indicate that the character of the wave functions for the VB hole is very different from that at the Fermi level in (GaMn)As.

We also performed similar experiments on other III-V-based FMS families (InGaMn)As and (InMn)As. From the viewpoint of both physics and applications, it is important to investigate the difference of the band structures of III-V based FMSs with different host materials. Such studies will give us a guideline on how to design heterostructures to accomplish intended functionalities. Also, it is useful to understand the whole band picture of FMSs. Especially, it is important to understand the (III,Mn)As (III=In and Ga) systems, which are the most prototypical FMSs. As the In content increases in InGaAs, the binding energy of Mn decreases from ~110 (GaAs) to ~30 meV (InAs).[71] This difference of binding

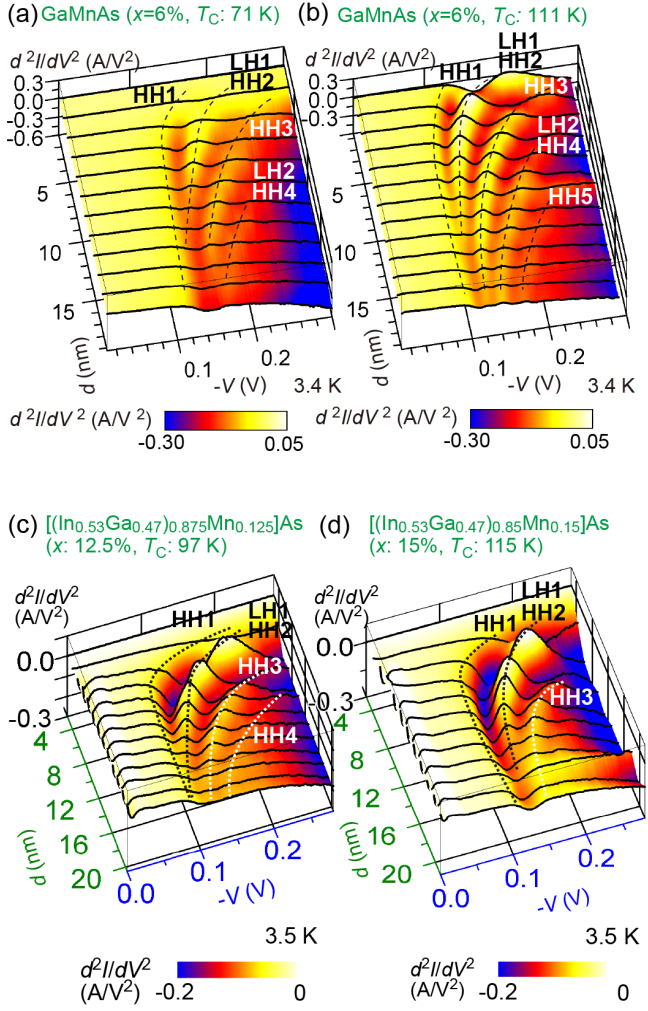

(a) GaMnAs (x=6%, $T_C$: 71 K)   (b) GaMnAs (x=6%, $T_C$: 111 K)

(c) [(In$_{0.53}$Ga$_{0.47}$)$_{0.875}$Mn$_{0.125}$]As (x: 12.5%, $T_C$: 97 K)

(d) [(In$_{0.53}$Ga$_{0.47}$)$_{0.85}$Mn$_{0.15}$]As (x: 15%, $T_C$: 115 K)

**FIG. 10.** (Color) The solid curves correspond to the $d^2I/dV^2$–$V$ curves observed in the device structure of Fig. 9: **(a)(b)** GaMnAs QW and **(c)(d)** (InGaMn)As QW with various thicknesses $d$ of (a) Ga$_{0.94}$Mn$_{0.06}$As ($T_C$ =71 K), (b) Ga$_{0.94}$Mn$_{0.06}$As ($T_C$ = 111 K), (c) [(In$_{0.53}$Ga$_{0.47}$)$_{0.875}$Mn$_{0.125}$]As ($T_C$ = 97 K), and (d) [(In$_{0.53}$Ga$_{0.47}$)$_{0.85}$Mn$_{0.15}$]As ($T_C$ = 115 K). The dashed curves trace the positions of the resonant peaks, where the assignment of these peaks is made on the basis of the calculated resonant levels. Colors in these graphs indicate the $d^2I/dV^2$ intensity extrapolated from the measured data. (a) and (b) are reprinted with permission from Nature Phys. **7**, 342 (2011). Copyright 2011 Nature Publishing Group. (c) and (d) are reprinted with permission from Phys. Rev. B **86**, 094418 (2012). Copyright 2012 American Physical Society.

energy may affect the way of IB formation or VB deformation by Mn doping. There have been conflicting reports about the understanding of the VB structure of InMnAs. For example, the cyclotron resonance experiments suggest the VB conduction in InMnAs,[72,73] while infrared optical absorption measurements on InMnAs show the double-exchange-like component, suggesting the IB conduction.[74] Thus, further systematic experiments are needed to investigate the VB structures of (InGaMn)As[75,77] and InMnAs.

The investigated heterostructures are composed of ([(In$_{0.53}$Ga$_{0.47}$)$_{1-x}$Mn$_x$]As or In$_{0.87}$Mn$_{0.13}$As) ($d$ nm) / AlAs (6 nm)/ In$_{0.53}$Ga$_{0.47}$As:Be (100 nm, Be concentration: $1 \times 10^{18}$ cm$^{-3}$) grown on $p^+$InP(001). In Figs. 10 (c) and (d), we show the examples of the $d^2I/dV^2$-$V$ curves (black solid curves) obtained for various $d$ of the [(In$_{0.53}$Ga$_{0.47}$)$_{1-x}$Mn$_x$]As RTDs with $x$=12.5% and 15%, respectively, where resonant oscillations are clearly observed in almost all the curves.[55] The oscillation-peak bias voltages get smaller, and the period of the oscillations become short as $d$ increases. Considering that the Fermi level corresponds to the zero bias condition, we conclude that the Fermi level exists in the band gap in (InGaMn)As. In the case of InMnAs diode (not shown here), we observed the HH1 level, although it was very weak due to the large strain in the InMnAs layer induced by the large lattice mismatch (~3%) between InP and InMnAs. Therefore, we can conclude that the Fermi level exists in the band gap in (III,Mn)As (III=In,Ga).

If there were large $p$-$d$ exchange splitting in VB, the resonant levels would be split and the splitting energy would increase with increasing $T_C$ or $x$. However, such behavior is not seen in Figs. 10(a)-(d), which means that the $p$-$d$ exchange splitting in VB is negligibly small in (III,Mn)As (III=In,Ga). In a RTD with a paramagnetic ZnMnSe QW, clear spin-splitting of the resonant level was observed, and this splitting energy increased with a magnetic field.[78] Our results mean that (III,Mn)As has a quite different band picture from that of II-VI based DMSs.[13]

With a multiband transfer matrix technique developed in Ref. 40 with the Luttinger-Kohn 6×6 **k·p** Hamiltonian and the strain Hamiltonian,[47] we calculated the quantum levels in these heterostructures. Here, we show the example of the calculation in the case of the $[(In_{0.53}Ga_{0.47})_{1-x}Mn_x]$As RTD with $x$=12.5%. To reproduce the experimental data of the resonant peak bias voltages $V_R$, we assumed the VB diagrams shown in Fig. 11 (a). Here, the thin green and thick orange lines express the LH and HH bands, respectively. The HH and LH bands are split especially in AlAs due to the large strain by the lattice mismatch (~4%) between AlAs and InP. Since the $p$-$d$ exchange splitting was negligibly small, we neglected the $p$-$d$ exchange interaction in our calculations. We divided the surface Schottky barrier region into three equally spaced regions and assumed that the band is flat in each region. $E_F$ is defined as the energy difference between the Fermi level and the HH band at the $\Gamma$ point in (InGaMn)As. We used $E_F$ and the deformation anisotropic term[79] $Q_\varepsilon$ as fitting parameters, which were determined so that the relationship between the calculated resonant tunneling energy $E_R$ (relative to the Fermi level) and measured $V_R$ becomes linear, that is, the $V_R$-$E_R$ relation is expressed by $V_R = sE_R$, where $s$ is a fitting parameter corresponding to the slope of $V_R$-$E_R$.[79,80,81]

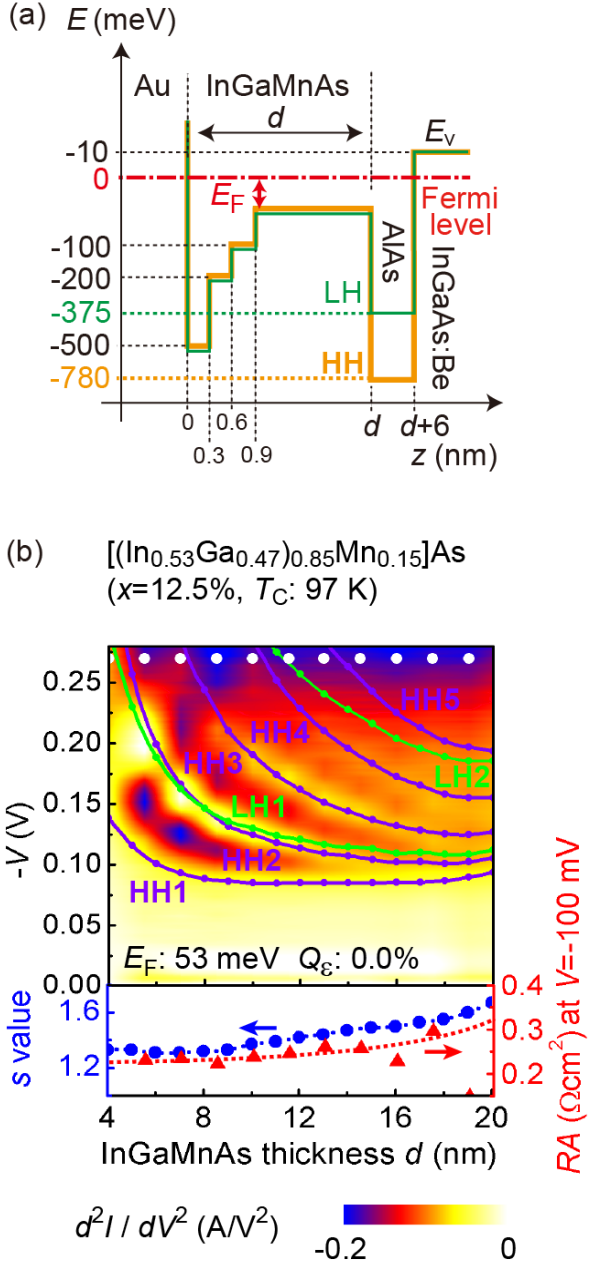

The color-coded $d^2I/dV^2$ intensity of the $[(In_{0.53}Ga_{0.47})_{1-x}Mn_x]$ RTD with $x$=12.5% as a function of $d$ is shown in the upper figure of Fig. 11 (b). Here, the intensity is extrapolated from the measured data at the $d$ values corresponding to the white dots shown at the top of these figures. The calculated $V_R$ values of the HH and LH resonant levels as a function of $d$ are expressed by the connected violet and green dots, respectively. The calculated HH and LH resonant levels show good agreement with the measured data. In the lower figure of Fig. 11 (b), blue and red dots express the $s$ value used in the fitting and $RA$, respectively. The $RA$ value changes as a function of $d$, which can be attributed to the diffusion of the interstitial Mn atoms to the surface. The interstitial Mn atoms work as donors and compensate holes near the surface in the thick *d* samples, leading to the increase in *RA*.[54] Therefore, we changed *s* gradually as *d* increases. We see a similar behavior between the *RA* - *d* and *s* - *d* curves, though their shapes are not perfectly the same. In our experiments, there is also a tunneling sequence where the holes injected in the VB quantum levels lose their energy and drop to IB by scattering in the (InGaMn)As QW, so the observed *RA* values include the contribution of the conduction of the IB holes, whereas *s* is determined just by the energy positions of the VB resonant levels. These probably cause the slightly different behavior between the *RA* - *d* and *s* - *d* curves. The resonant peaks observed in all of our samples containing (GaMn)As and (InGaMn)As QWs are well reproduced by our model, which supports our conclusion that the resonant levels are formed by the quantization of the VB holes. In the case of InMnAs, although HH1 was observed, higher resonant levels (LH1 and HH3) were not clearly observed. The reason for this result is not clear, but it is probably due to the degradation of the crystallinity of InMnAs induced by the strong strain. Even in the case of the InMnAs diode, however, we roughly reproduced the experimental resonant peaks by using our model. For more systematic and precise analysis, strain-free heterostructures are necessary for InMnAs, although it is very difficult to achieve them using typical III-V materials.

FIG. 11. (Color) **(a)** VB diagram of the (InGaMn)As resonant tunneling device assumed in our calculation. Here, the thin green and thick orange lines express the VB of the LH and HH bands at the $\Gamma$ point, respectively. The red dash-dotted line corresponds to the Fermi level. **(b)** The upper graph show the comparison between the calculated resonant levels and the experimentally obtained $d^2I/dV^2$ data of $[(In_{0.53}Ga_{0.47})_{0.875}Mn_{0.125}]$As, as functions of $-V$ and $d$. The $d^2I/dV^2$ intensity is expressed by colors, whose intensities are extrapolated from the measured data with $d$ corresponding to the white dots shown at the top of these figures. The connected violet and green dots are the calculated resonant peak bias voltages $V_R$ of the HH and LH bands, respectively. In the lower graph, blue and red dots express the $s$ value used in the fitting and the resistance area product $RA$, respectively. Reprinted with permission from Phys. Rev. B **86**, 094418 (2012). Copyright 2012 American Physical Society.

Here, we show that one can rule out the possibility that these oscillations observed in our studies are induced by quantized two-dimensional hole-gas states that might be formed at the interface between the AlAs and GaAs:Be (or InGaAs:Be).[82,83] We point out that the results observed in our experiments show typical features of the QW thickness dependence of resonant tunneling. For example, these oscillations clearly tend to be weakened with increasing *d*, which can be understood by weakening of the quantization. Other essential feature is that the VB gradually branches out to sub-band levels with decreasing *d*. In Fig. 10, HH1 and LH1 (or HH2) levels are nearly merged at around *d* = 14 - 16 nm, but these levels branch out with decreasing *d*. These features are quite different from those of the quantized two-dimensional hole-gas states at AlAs/GaAs:Be (or InGaAs:Be), if any. Also, if the oscillations were induced by quantized two-dimensional hole-gas states, the oscillation peak bias voltage would be proportional to *RA* as mentioned in Ref. 82. However, as can be seen in Fig. 11 (b), all the oscillation peak voltages become smaller even when *RA* increases with increasing *d*. Furthermore, as shown in Ref. 83, we did not see any clear oscillations induced by resonant tunneling in the $d^2I/dV^2$-*V* curves of the tunnel device composed of GaMnAs/ AlAs (5nm)/ GaAs:Be ($1\times10^{18}$ cm$^{-3}$) when the surface GaMnAs thickness was as thick as 30 nm, where the quantization of the holes in GaMnAs is very weak. If the $d^2I/dV^2$ oscillations observed in our study were induced by the quantized two-dimensional hole-gas states, the oscillation would be observed even when the GaMnAs thickness was 30 nm. This result means that it is impossible to detect the quantized two-dimensional hole-gas states in our experimental setup when the Be doping level is $\sim1\times10^{18}$ cm$^{-3}$.

The quantized two-dimensional hole-gas states formed at the AlAs/GaAs:Be interface have been observed when the Be concentration is extremely low (for example, $6\times10^{14}$ cm$^{-3}$ in Ref. 84). As the Be concentration increases, however, the quantization of these states becomes weak. Also, as shown in Ref. 82, the energy separation between these states is only several meV when the Be concentration is $1\times10^{18}$ cm$^{-3}$. These are the reasons why we did not detect any quantized two-dimensional hole-gas states in our experiments. Also, we do not see any clear signals related to these states at the AlAs/GaAs:Be(or a thin GaAs spacer) interface in the studies of the similar p-type GaAs-based resonant tunneling structures with the Be concentration of as high as ~$10^{18}$ cm$^{-3}$ [80,81,85]. Therefore, we conclude that the oscillations observed in our study are attributed to the resonant tunneling in the FMS QW layers.

The disadvantage of our method is that it is difficult to identify the signal coming from IB, and thus we were not able to obtain any clear information about IB in our studies. However, we note that more recent experiments using resonant angle-resolved photoemission spectroscopy (ARPES) have succeeded in detecting IB in GaMnAs; Kobayashi *et al.* have investigated GaMnAs films with various Mn concentrations (2-10%) and found that the Fermi level exists above VB in all the GaMnAs samples.[86] Furthermore, non-dispersive IB was detected near the Fermi level, overlapping with VB over a wide range of the energy region from 400 meV below the VB top up to the Fermi level. It was also found that the HH and LH bands of (GaMn)As are almost the same as those of GaAs even with such an overlap with IB, which is completely consistent with our resonant tunneling experiments. Also, Fujii *et al.* have carried out bulk-sensitive ARPES measurements and showed that the Fermi level exists 50 meV above the host VB top position in Ga$_{0.87}$Mn$_{0.13}$As, which is quantitatively consistent with our studies.[87] We note that there are a lot of papers supporting the IB conduction picture by using various methods recently.[88,89,90,91,92,93]

In Fig. 12(a), the triangular red, circular blue, and rectangular green points show the $E_F$ values of Ga$_{1-x}$Mn$_x$As, [(In$_{0.53}$Ga$_{0.47}$)$_{1-x}$Mn$_x$]As, and In$_{0.87}$Mn$_{0.13}$As obtained in our study, respectively, where the values next to the data points are the $T_C$ values. We see that $E_F$ decreases as the In content increases, which can be attributed to the reduction of the binding energy of the Mn states. In both GaMnAs and (InGaMn)As, $E_F$ decreases as $T_C$ increases when $x$ is fixed, which can be explained by the increase in the IB hole concentration as shown in Fig. 12(b). Also, $E_F$ increases as $x$ increases, which can be qualitatively explained by the IB broadening as shown in Fig. 12(c). Similar behavior was theoretically expected in Refs. 65 and 94.

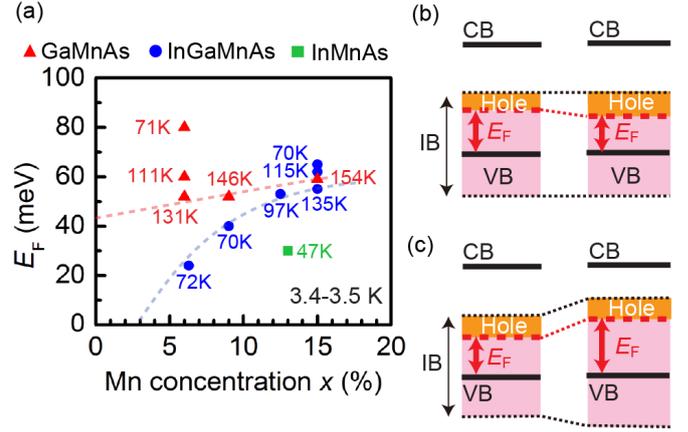

FIG. 12. (Color) **(a)** Triangular red, circular blue, and rectangular green dots are the $E_F$ values of GaMnAs, (InGaMn)As, and InMnAs respectively. The values next to the data points in this figure are the $T_C$ values. **(b)** Schematic band pictures showing how $E_F$ decreases with increasing $T_C$ or hole concentration when $x$ is fixed. **(c)** Schematic band pictures showing how $E_F$ increases with increasing $x$. Here, CB means the conduction band. Reprinted with permission from Phys. Rev. B **86**, 094418 (2012). Copyright 2012 American Physical Society.

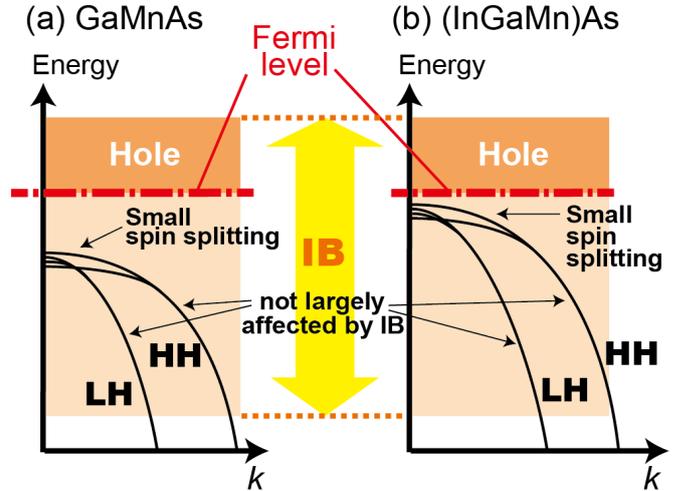

FIG. 13. (Color) The schematic band pictures of **(a)** GaMnAs and **(b)** (InGaMn)As (including (InMn)As) obtained by our studies combined with the recent angle resolved photoemission spectroscopy measurements, where the IB is expressed by orange areas and the Fermi level is expressed red dashed-dotted line. IB is overlapped on the VB, but VB structure itself is not largely affected (the effective mass is not changed.) The *p-d* exchange splitting is small and is estimated to be less than ~5 meV for LH.

From these results, we can depict the band structure picture of (GaMn)As and (InGaMn)As as shown in Figs. 13 (a) and (b), respectively. Even though the VB and IB are largely overlapped, the VB ordering is not much affected by the IB. The transport is dominated by the IB holes, not by the VB holes. As the In content increases, the Fermi level becomes close to the VB top, probably because of the band gap reduction.

## 4. Mn concentration dependence of the Fermi level position in GaMnAs[95]

Investigating the Mn concentration dependence of the Fermi level gives us important information on the mechanism of the ferromagnetism in Ga$_{1-x}$Mn$_x$As.[95] Especially, metal-insulator transition (MIT), which occurs at $x$ = 1 – 2 %, has not been well understood in GaMnAs. Previously, the VB conduction picture has been widely accepted in GaMnAs,[47,56] where the MIT of GaMnAs was understood by the Fermi level crossing over the VB[96,97], similarly to $p$-type GaAs doped with non-magnetic acceptors such as Be or Zn. The ferromagnetism in GaMnAs has been thought to be induced by the VB holes interacting with the localized $d$ electrons of the Mn atoms.[47,56, 98] However, recently, many experiments have shown the strong evidence that the Fermi level exists in IB in the band gap, which requires reconsideration on the above scenario. Thus, to clarify the origin of the ferromagnetism and MIT in GaMnAs, it is essential to precisely investigate the Fermi level position and the VB structure of GaMnAs in the low Mn content region including the onset of ferromagnetism and MIT.

We carefully investigated the VB structure and the Fermi level position in a series of Ga$_{1-x}$Mn$_x$As from the insulating region ($x$=~0.01%) to the metallic region ($x$=3.2%) by using the resonant tunneling spectroscopy in the DB-QW heterostructures composed of (from the top to the bottom) Ga$_{0.94}$Mn$_{0.06}$As (20 nm) / AlAs (6 nm) / Ga$_{1-x}$Mn$_x$As QW ($d$ nm) / AlAs (6 nm) / GaAs:Be

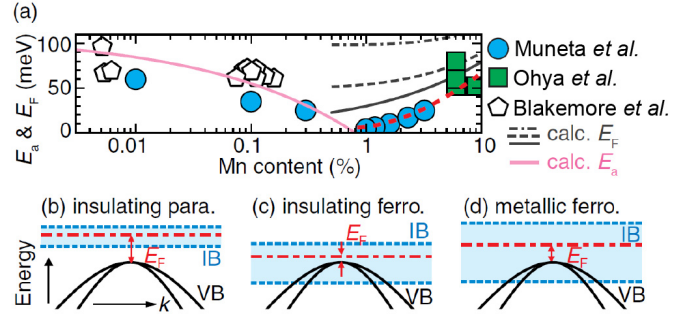

FIG. 14. (Color) (a) The blue circles and green squares are the $E_F$ values obtained in our studies. The blank pentagons are the $E_a$ values reported in Ref. 100. The pink curve is the calculated $E_a$ obtained from the equation $0.11[1-(x/0.8)^{1/3}]$. The gray dash-dotted, dashed, and solid curves are the calculated $E_F$ with respect to $E_V$ by the VBAC model[99] and the free-electron approximation. In the present VBAC model calculation, $E_{Imp}$ is assumed to be 0.1, 0.05, and 0.01 eV, respectively. The red dashed curve connects the $E_F$ values in $x$ > 1%. (b)-(d) The VB and IB diagrams of GaMnAs derived from this study in the (b) insulating paramagnetic ($x$ < 1%), (c) insulating ferromagnetic ($x$ = 1 - 2%), and (d) metallic ferromagnetic ($x$ > 2%) regions. The black solid curves are the VB. The blue dotted lines are the upper and bottom edges of the IB. The blue region represents the IB region. The red dash-dotted lines are the Fermi level. Reprinted with permission from Appl. Phys. Lett. **103**, 032411 (2013). Copyright 2013 American Institute of Physics.

(100 nm, Be concentration: 1×10$^{18}$ cm$^{-3}$). To fabricate the RTDs with various GaMnAs-QW thickness $d$ on a wafer, we linearly moved a shutter in front of the substrate while growing the Ga$_{1-x}$Mn$_x$As QW, in which $d$ was varied from 10 nm to 16 nm within the wafer of 15×10 mm.[2,26] We observed clear resonant tunneling peaks in all the samples, which shifted and tended to converge at certain negative bias voltages as $d$ increases.

Using the same theoretical calculation mentioned above, we estimated the $E_F$ values in these samples. In Fig.14 (a), the green squares and blue circles are the $E_F$ values obtained in our studies of Ref. 54 and 95, respectively. The black pentagons are the thermal-activation energy $E_a$ of the Mn acceptors in GaAs obtained by the magneto-transport measurements in Ref. 100. We see anomalous behavior that the Fermi level of (Ga$_{1-x}$Mn$_x$)As becomes closest to the VB top at $x$=1.0% corresponding to the onset of ferromagnetism near MIT, but it moves away from the VB as $x$ increases or decreases from 1.0%.

The pink curve is the calculated $E_a$ values obtained from the equation $0.11[1-(x/0.8)^{1/3}]$ mentioned in Ref. [100]. Here, we selected 0.8% as the intercept of $x$ to fit the curve to the experimental data. The gray dash-dotted, dashed, and solid curves are the calculated $E_F$ by the valence-band anti-crossing (VBAC) model[99] with the free-electron approximation. In the VBAC, the impurity level $E_{Imp}$ is assumed to be 0.1, 0.05, and 0.01 eV, respectively. Also, the anti-crossing coupling constant $C_{Mn}$ is assumed to be 0.18 eV. The free-electron approximation calculation for the Fermi level position in the IB is done by assuming the hole concentration $p$ to be $x/2$ and the effective mass to be $10m_0$,[61] where, $m_0$ is the free electron mass. Figures 14 (b)-(d) show the VB and IB diagrams derived from our study in the (b) insulating paramagnetic ($x < 1\%$), (c) insulating ferromagnetic ($x = 1 - 2\%$), and (d) metallic ferromagnetic ($x > 2\%$) regions, respectively. The black solid curves correspond to the VB. The blue dotted lines correspond to the upper and bottom edges of the IB, the blue region to the IB region, and the red dash-dotted lines to the Fermi level. In the paramagnetic GaMnAs ($x < 1\%$), $E_F$ decreases as $x$ increases, which is quantitatively consistent with the activation-energy-lowering effect observed in the magneto-transport measurements.[100] This behavior is the same as that observed in the insulating region of the non-magnetic accepter-doped $p$-type GaAs.[101] The Fermi level behavior in the paramagnetic region is caused by the screening effect due to the heavy Mn doping, which makes the IB position close to the VB. Meanwhile, $E_F$ increases as $x$ increases in the ferromagnetic GaMnAs with $x >$ 1%, which means that the Fermi level moves away from the VB. This anomalous behavior is qualitatively explained by the VBAC model [the gray solid curve in Fig. 14 (a)]. The quantitative discrepancy between the experimental and calculated Fermi level is probably because this calculation does not take into account the screening and many body effects. This Fermi level behavior in the ferromagnetic region means that the IB truly exists in the ferromagnetic GaMnAs and the anti-crossing interaction is induced by the electron-electron interaction.[65]

Figure 14 (a) shows that the Fermi level exists in the IB in the band gap in the whole $x$ region, which suggests that the ferromagnetism of GaMnAs is strongly related to the IB. At the MIT border ($x=\sim 2\%$), the Fermi level is still in the band gap, which suggests that the MIT occurs in the IB. This behavior of the Fermi level is completely different from that in the case of the heavily-doped $p$-type GaAs with the non-magnetic acceptors and contradicts the VB conduction picture, where the IB completely merges into the VB. The IB still exists in the metallic GaMnAs, which means that the impurity states still remain in the band gap. This is probably because the holes are trapped in the acceptor states induced by $p$-$d$ hybridized orbitals [102, 103] after the Coulomb potential is completely screened. At the onset of the ferromagnetism ($x = \sim 1\%$), the $x$ dependence of the Fermi level changes, which is thought to be induced by the change of the dominant effect for determining the IB position from the screening effect to the anti-crossing interaction. This must give a clue to understanding the mechanism of the ferromagnetism. We note that this anomalous behavior of the Fermi level is not directly related to the MIT because the MIT and the turning up of the Fermi level position occur at slightly different values of $x$ (1.5 - 2% and ~1%, respectively).

## C. Potential applications of GaMnAs hetero-structures

### 1. Magnetic tunnel transistor (MTT)[104]

GaMnAs-based three-terminal devices can be a good model system for future spintronic devices with power amplifiability; they can be applied to integrated circuits with various functions. As one of the three-terminal spintronic devices, metal-based magnetic tunnel transistors (MTTs) have generated much attention.[105,106,107] Metal-based MTTs are composed of ferromagnet (FM) / insulator (I) / FM / semiconductor (SC) heterostructures, whose output currents can be controlled by the bias voltages and magnetization orientation of the FM layers. However, it is very difficult to fabricate metal-based MTTs with amplifiability, because the transfer ratio $\alpha$ defined by $I_C / I_E$ ($I_C$ and $I_E$ are the collector and emitter current, respectively) is very low ($\leq 10^{-2}$) due to the frequent scatterings in the metallic base.

We have proposed a semiconductor spin hot-carrier transistor (SSHCT)[104]. SSHCT is structurally similar to the conventional metal-based MTTs but composed of all-epitaxial semiconductor heterostructures, which enables advanced band engineering and has good compatibility with the existing semiconductor technology. We fabricated a three-terminal device composed of $Ga_{0.95}Mn_{0.05}As$ (Emitter: 30 nm)/ GaAs (1 nm)/ AlAs (2 nm)/ GaAs (1 nm)/ $Ga_{0.95}Mn_{0.05}As$ (Base: 30 nm)/ Be-doped GaAs (30 nm) on $p$-type GaAs(001) substrates (Collector). A part of the top GaMnAs layer was etched and a contact was made on the GaMnAs base layer. We have successfully controlled the potential of the GaMnAs base layer by applying the voltage. The emitter-base bias voltage $V_{EB}$ dependence of the collector current $I_C$, emitter current $I_E$, and base current $I_B$ showed that the current transfer ratio $\alpha$ (= $I_C / I_E$) and the current gain $\beta$ (= $I_C / I_B$) are 0.8 - 0.95 and 1 - 10, respectively, which means that GaMnAs-based SSHCTs have current amplification capability. Unfortunately, because the leakage current through the low-temperature grown GaAs barrier between the collector and base was not sufficiently suppressed, we were not able to obtain power amplification capability. If the barrier quality can be improved, high-performance SSHCT will be achieved.

## 2. Three-terminal RTD device with a GaMnAs QW[108]

"Three-terminal" GaMnAs QW-DB devices such as spin resonant-tunneling transistors, where the spin-dependent quantum levels can be externally controlled by applying the voltage to the electrode connected to the GaMnAs QW, are very attractive. In comparison with the metal-based MTTs, we can more effectively design the heterostructure for controlling the spin-polarized coherent holes utilizing the well-established band-engineering technique of III-V semiconductors. Furthermore, the metallic conductivity of GaMnAs allows us to make a good contact to the GaMnAs QW layer, and thus more efficient control of the quantum levels is possible than that in usual semiconductor-based three-terminal quantum heterostructures containing a QW electrode. [109] We fabricated a RTD composed of $Ga_{0.94}Mn_{0.06}As$(10 nm)/ GaAs (1 nm)/ $Al_{0.94}Mn_{0.06}As$(4 nm)/ GaAs(2 nm)/ $Ga_{0.94}Mn_{0.06}As$ QW(2.5 nm)/ GaAs(1 nm)/ AlAs(4 nm)/ GaAs:Be (100 nm) on a $p^+$GaAs(001) with a "third" electrode to modulate the potential of the GaMnAs QW. We successfully modulated the resonant levels in the GaMnAs QW and the magnetocurrent ratio by modulating the voltage of the third electrode, showing that it is possible to access the potential of the GaMnAs QW from the outside electrode[108]. This technique can be applied to the quantum spin devices, such as spin resonant-tunneling transistors and magnetic monostable-bistable transition logic elements.[110]

## 3. Double-quantum well heterostructure with a GaMnAs QW and a GaAs QW[111]

Ferromagnetic double-quantum-well (DQW) heterostructures are also potentially promising, because we can expect sharp spin-filtering due to the strong resonant tunneling in the structures. In ferromagnetic GaMnAs and non-magnetic GaAs multi-QW RTDs, a strong spin filter effect and spin polarization oscillation have been predicted theoretically.[112, 113] The DQW-RTD structure used for the device is composed of (from the top to the bottom) $Ga_{0.94}Mn_{0.06}As$ (10 nm) / GaAs (1 nm) / AlAs (1.5 nm) / GaAs (2 nm) / $Ga_{0.94}Mn_{0.06}As$ QW (2.5 nm) / GaAs (1 nm) / AlAs (4 nm) / GaAs QW (4 nm) / AlAs (4 nm) / GaAs:Be (100 nm) grown on a $p^+$GaAs (001) substrate. After the growth, we fabricated 24-μm-diameter mesas by photolithography and conventional chemical wet etching. We observed a clear resonant tunneling effect due to the quantum confinement in the undoped GaAs QW, which results in negative differential resistance at the resonant peak of LH1 in the GaAs QW. Also, we see a slight shift of the resonant peaks between parallel and antiparallel magnetization, which is induced by the TMR of the top GaMnAs / AlAs / GaMnAs QW MTJ. The shift of the resonant peaks and the resonant tunneling characteristics of the non-magnetic GaAs QW have led to the TMR oscillation as a function of the bias voltage. We observed negative TMR at the resonant peak of LH1 in the negative bias region (where holes are injected from the GaAs:Be electrode), which is due to the negative differential resistance and the shift of the resonant peaks bias voltages between parallel and antiparallel magnetization. Our results indicate that the TMR can be artificially controlled by utilizing the sharp resonant levels in the GaAs QW without using any other devices outside,[114] which will open up a new possibility for controlling the spin signal in the ferromagnetic heterostructures.

## III. PROPERTIES OF N-TYPE (InFe)As

### A. Necessity of n-type ferromagnetic semiconductors

Ferromagnetic semiconductors (FMSs) with carrier-induced ferromagnetism have been intensively studied for decades as they have novel functionalities that cannot be achieved with conventional metallic materials. These include the ability to control magnetism by electrical gating[115] or light irradiation,[116] while fully inheriting the advantages of semiconductor materials such as band engineering.[117] Lots of studies on FMSs have been concentrated on III-V semiconductors doped with Mn, such as (In,Mn)As[15,17,18] and (Ga,Mn)As, [19,20, 118], which are always p-type with hole densities as high as $10^{20}$ - $10^{21}$ cm$^{-3}$. In those materials, Mn atoms work not only as local magnetic moments but also as acceptors providing holes that mediate ferromagnetism. This behavior, however, brings about a severe drawback; it is difficult to control the ferromagnetism and carrier type (in other words, Fermi level) independently. This problem makes it difficult to utilize the Mn-based FMSs for practical devices, as well as to understand the mechanism of carrier-mediated ferromagnetism in which controlling the Fermi level is very important.

All of semiconductor devices, including pn junction diodes, field-effect transistors, and semiconductor lasers, require a pair of n-type and p-type semiconductor materials to work. Semiconductor spintronics devices are no exception. Despite the extensive studies on magnetic semiconductors, reliable n-type carrier-induced FMSs are still missing. In the following sections, we show that by introducing iron (Fe) atoms into InAs, it is possible to fabricate a new FMS with the ability to control ferromagnetism by both Fe and independent carrier doping.[120,121] Despite the general belief that the tetrahedral Fe-As bonding is antiferromagnetic, we demonstrate that (In,Fe)As doped with electrons behaves as an n-type electron-induced FMS, a missing piece of semiconductor spintronics for decades. This opens the way to realize novel spin-devices such as spin light-emitting diodes or spin field-effect transistors, as well as helps understand the mechanism of carrier-mediated ferromagnetism in FMSs.

In the following sections, we describe the MBE growth, structural, transport, magneto-optical, magnetization, and magneto-transport properties of (In,Fe)As. We also discuss the bandstructure, Fermi level, effective mass of electrons, and large s-d exchange interaction of this material.

## B. Molecular-beam epitaxy growth and structural characterizations

We have grown 100-nm-thick $(In_{1-x},Fe_x)As$ films by low-temperature molecular-beam epitaxy (LT-MBE) on semi-insulating GaAs substrates, as shown in Fig. 15 (a). After growing a 50 nm-thick GaAs buffer layer at 580°C, we grew a 10 ~ 20 nm-thick InAs buffer layer at 500°C. The growth of InAs at high temperature helps to quickly relax the lattice mismatch between InAs and GaAs, and create a relatively smooth InAs surface. After cooling the substrate temperature to 236°C, we started to grow a 100 nm–thick (In,Fe)As with or without Be co-doping. Finally, we grew a 5 ~ 10 nm InAs cap (except for sample B0 with a 20 nm cap) to prevent oxidation of the underlying (In,Fe)As layer. At this low growth temperature, we found that doped Be atoms act as donors rather than acceptors. Two series of $(In_{1-x},Fe_x)As$ samples were grown as summarized in Table I.

**Table I.** List of $(In_{1-x},Fe_x)As$ samples studied in this work.

| Sample | Fe concentration $x$ (%) | Electron concentration $n$ (cm$^{-3}$) | Non-magnetic dopants |
|--------|------|----------------------|------|
| A1 | 5.0 | $1.8 \times 10^{18}$ | Be |
| A2 | 5.0 | $2.9 \times 10^{18}$ | Be |
| A3 | 5.0 | $6.2 \times 10^{18}$ | Be |
| A4 | 5.0 | $1.8 \times 10^{19}$ | Be |
| B0 | 9.1 | $1.6 \times 10^{18}$ | None |
| B1 | 8.0 | $1.3 \times 10^{18}$ | Be |
| B2 | 8.0 | $1.5 \times 10^{18}$ | Be |
| B3 | 8.0 | $9.4 \times 10^{18}$ | Be |
| B4 | 8.0 | $2.8 \times 10^{19}$ | Be |

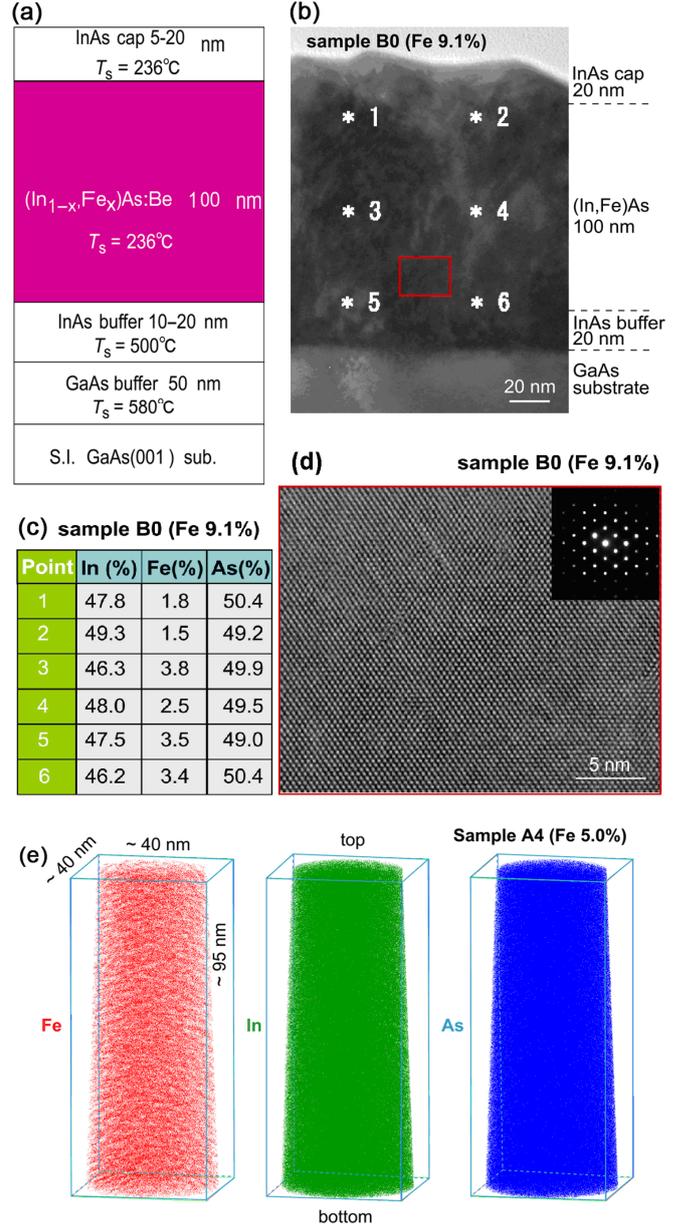

**FIG. 15.** (Color) **(a)** Schematic sample structure studied in this work. **(b)** Transmission electron microscopy (TEM) image of a 100 nm-thick $(In_{0.909},Fe_{0.091})As$ layer (sample B0 in table I) grown on a GaAs substrate, taken along the GaAs[110] axis. **(c)** In, Fe and As atomic concentrations obtained by energy dispersive X-ray spectroscopy (EDX) taken at 6 points marked by * in the above TEM image. It is observed that the As atomic concentration (~50%) is close to the sum of the In and Fe atomic concentrations, indicating that Fe mostly reside at the In site, although there are fluctuations of Fe concentration depending on the location. **(d)** High-resolution TEM (HRTEM) lattice image taken at an $(In_{0.909},Fe_{0.091})As$ area close to the substrate (marked by the red rectangular in (b)). Inset shows the transmission electron diffraction of this (In,Fe)As layer. The (In,Fe)As layer shows zinc-blende crystal structure only. Other HRTEM images taken at areas close to the surface and in the middle of this (In,Fe)As layer show no inter-metallic precipitation, although there are some stacking faults due to the large lattice mismatch between the (In,Fe)As layer and the GaAs

substrate. (e) Three-dimensional atom distribution of Fe, In and As in a 100 nm-thick $(In_{0.95},Fe_{0.05})As$ layer (sample A4), obtained by the laser assisted three-dimensional atom probe (3DAP) technique. One dot (red, green, blue) corresponds to one (Fe, In, As) atom. Reprinted with permission from Appl. Phys. Lett. **101**, 182403 (2012). Copyright 2012 American Institute of Physics.

Series A with a Fe concentration of $x = 5.0\%$ and series B with a higher Fe concentration of $x = 8.0\%$ (except for B0 with $x = 9.1\%$) were grown at a substrate temperature of 236°C, with and without electron doping.

Fig. 15(b) shows a transmission electron microscopy (TEM) image of sample B0, which is $(In_{0.909},Fe_{0.091})As$ without Be co-doping. Fig. 15 (c) shows the In, Fe and As atomic concentrations at six different points (*1 - *6) in sample B0 shown in Fig. 15 (b), obtained by energy dispersive x-ray (EDX) spectroscopy. It is observed that the As atomic concentration is nearly equal to the sum of the In and Fe atomic concentrations, revealing that most of the Fe atoms reside at the In sites. Fig. 15 (d) shows a high-resolution TEM lattice image of an area close to the buffer layer, indicated by the red rectangular in Fig. 15 (b). The inset in Fig. 15 (d) shows a transmission electron diffraction (TED) pattern of this (In,Fe)As layer. Despite low-temperature growth, the whole (In,Fe)As layer shows zinc-blende crystal structure and no visible metallic Fe or inter-metallic Fe-As precipitation. We further thinned the TEM specimen down to ∼ 10 nm and found no evidence of such metallic Fe or inter-metallic Fe-As precipitated particles, demonstrating that it is possible to grow zinc-blende (In,Fe)As of good quality by LT-MBE. Similar TEM images of samples A4 and B4, which are both Be-doped, show that Be co-doping does not affect the zinc-blende structure of (In,Fe)As. Furthermore, we have used the laser-assisted three-dimensional atom probe (3DAP) technique[122] to investigate the distribution of Fe, In and As in (In,Fe)As with a nearly atomic resolution. Fig. 15 (e) shows the three-dimensional atom distributions of In, Fe and As in sample A4, obtained by 3DAP. It can be seen that the Fe atoms distribute everywhere in the (In,Fe)As with local fluctuation of Fe concentration. In order to find any precipitation of metallic Fe or inter-metallic Fe-As nanoclusters, we divided the observed area into about 18000 blocks (200 atoms/block) and investigated the local Fe, In and As atomic distributions in those blocks. We found that there is no block with metallic Fe nanoclusters (Fe 100%, In 0%, As 0%) or inter-metallic / zinc-blende type Fe-As nanoclusters (In 0%). Atomic distributions of In, Fe and As in all blocks can be described by $(In_{1-x}Fe_x)_{0.5}As_{0.5}$. These analyses together with TEM characterizations indicate that the (In,Fe)As layer is composed of a single-phase zinc-blende crystal, and the Fe concentration fluctuation occurs only within the group-III sublattice without precipitation of any other phase.

## C. Doping and electrical properties

At the In sites, the Fe ions have two possible states; acceptor state ($Fe^{2+}$) and neutral state ($Fe^{3+}$). If the $Fe^{2+}$ states were dominant, (In,Fe)As layers would be p-type and the hole concentration would be close to the doped Fe concentration at room temperature, similar to the case of (In,Mn)As. In reality, however, sample B0 (and all the other undoped samples) shows n-type with a maximum residual electron concentration of $1.8 \times 10^{18}$ cm$^{-3}$ at room temperature, which is four orders of magnitude smaller than the doped Fe concentration.

We measured temperature dependence of the electron mobility of (In,Fe)As, and the result indicates that the neutral impurity scattering, rather than the ionized impurity scattering, is dominant up to room temperature. Fig. 16 shows the temperature dependence of the electron mobility $\mu$ of the as-grown $(In_{0.909},Fe_{0.091})As$ (sample B0), with vertical and horizontal axes plotted in the logarithmic scale. The dashed red line is the fitting with $\mu \sim T^\gamma$. For $T < 50$ K, the mobility is nearly temperature-independent ($\gamma = 0.04$). For $T > 50$ K, the mobility weakly depends on temperature ($\gamma = 0.25$). This suggests that the Fe impurities in this material remain neutral. If the Fe impurities were ionized (i.e. in the acceptor $Fe^{2+}$ state), the sample would be p-type and the $\mu - T$ relation would be given by

$$\mu \sim \frac{1}{n_{ion}}(2k_BT)^{3/2}[\ln\left(1+\alpha k_B^2T^2\right)]$$

for ionized impurity scattering, requiring $\gamma \geq 1.5$. In reality, however, the sample is n-type and $\gamma$ is close to zero, indicating that the Fe impurities remain in the neutral state. When the neutral impurity scattering dominates, the $\mu - T$ relation is given by

$$\mu \sim \frac{1}{n_{neutral}} \times const ,$$

which is nearly temperature-independent. (Here, we neglect the contribution of phonon scattering, which is negligible at low temperatures. The contribution of alloy scattering cannot be a dominant scattering mechanism since it would give a negative $\gamma$.) The electron mobility of sample B0 and other (In,Fe)As samples is several tens of cm²/Vs, which is an order of magnitude higher than the hole mobility of (Ga,Mn)As. All of these facts indicate that the Fe atoms in (In,Fe)As are in the neutral state ($Fe^{3+}$) rather than in the acceptor state ($Fe^{2+}$). This result is consistent with the chemical trend of Fe in other III-V semiconductors,[123] and the results of electron paramagnetic resonance of Fe impurity in InAs, which shows the isoelectronic $Fe^{3+}$ state with $3d^5$ configuration (5 $\mu_B$ / Fe atom).[124] The present result is also similar to that obtained in the previous work[125] on paramagnetic (Ga,Fe)As, in which Fe atoms were found to reside at the Ga sites and in the $Fe^{3+}$ state. The residual electrons in sample B0 probably come from the As anti-site defects acting as donors due to the LT-MBE growth.[126] Since the Fe impurities contribute to spin (magnetization) but not to carrier generation, we have an important degree of freedom in controlling the carrier type and carrier concentration by independent chemical doping.

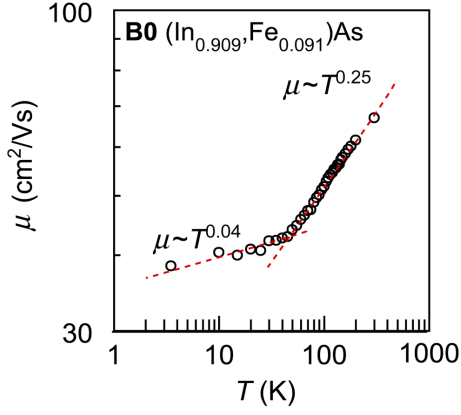

**FIG. 16.** (Color online) Temperature dependence of the electron mobility of sample B0, which indicates the neutral state of the Fe impurities on In sites in (In,Fe)As. Dashed red line is the fitting $\mu \sim T^\gamma$. Reprinted with permission from Appl. Phys. Lett. **101**, 182403 (2012). Copyright 2012 American Institute of Physics.

We then tried doping (In,Fe)As layers with donors to see the carrier-induced ferromagnetism. After trying several doping methods, we found that Beryllium (Be) atoms doped in (In,Fe)As at a low growth temperature of $T_S = 236$°C work as good double donors, not as acceptors as in the case of Be-doped InAs grown at $T_S > 400$°C. This suggests that Be atoms reside at interstitial positions due to the low growth temperature. For these electron doped (In,Fe)As layers, we investigate their ferromagnetism by using magnetic circular dichroism (MCD), superconducting quantum interference device (SQUID), and anomalous Hall effect (AHE) measurements. Despite the general belief that the tetrahedral Fe-As bonding is antiferromagnetic,[127] all of our data show appearance and evolution of ferromagnetism in (In,Fe)As with increasing both the Fe concentration ($x = 5 - 8\%$) and electron concentration ($n = 1.8 \times 10^{18}$ cm⁻³ - $2.7 \times 10^{19}$ cm⁻³), indicating that (In,Fe)As is an intrinsic n-type FMS, and that we can control the ferromagnetism of this material independently by Fe doping and electron doping.[120,121]

## D. Magnetic circular dichroism (MCD) and bandstructure

MCD is a technique that measures the difference between the reflectivity of right ($R_{\sigma+}$) and left ($R_{\sigma-}$) circular polarisations, and its magnitude is given by

$$\text{MCD} = \frac{90}{\pi} \frac{(R_{\sigma+} - R_{\sigma-})}{2} \sim \frac{90}{\pi} \frac{dR}{dE} \Delta E,$$

where $R$ is the reflectivity, $E$ is the photon energy, and $\Delta E$ is the spin-splitting energy (Zeeman energy) of a material. Since the MCD spectrum of a FMS directly probes its spin-polarized band structure induced by the s,p-d exchange interactions and its magnitude is proportional to the magnetization ($\Delta E \propto M$), MCD is a powerful and decisive tool to judge whether a FMS is intrinsic or not.[128] Note that the spectral features (i.e. enhanced at optical critical point energies of the host semiconductor), as well as the absolute magnitude of MCD, are important to judge whether a FMS is intrinsic or not.[129] Figures 17 (a) – (h) show the MCD spectra of sample series A (A1 - A4) and sample series B (B1 - B4), measured at 10 K under a magnetic field of 1 Tesla applied perpendicular to the film plane. With increasing the electron concentration $n$ and Fe concentration, the MCD intensity shows strong enhancement at optical critical point energies $E_1$ (2.61 eV), $E_1 + \Delta_1$ (2.88 eV), $E_0$' (4.39 eV) and $E_2$ (4.74 eV) of InAs, which show the magnetic "fingerprints" of (In,Fe)As. For sample B4, (In$_{0.92}$,Fe$_{0.08}$)As with $n = 2.8 \times 10^{19}$, the MCD peak at $E_1$ already reaches 100 mdeg at 10 K, which is two orders of magnitude greater than the MCD caused by the Zeeman splitting of InAs (~1 mdeg/Tesla)[128]. Therefore, the effective magnetic field acts on the InAs matrix is as large as 100 Tesla; thus, it cannot be explained

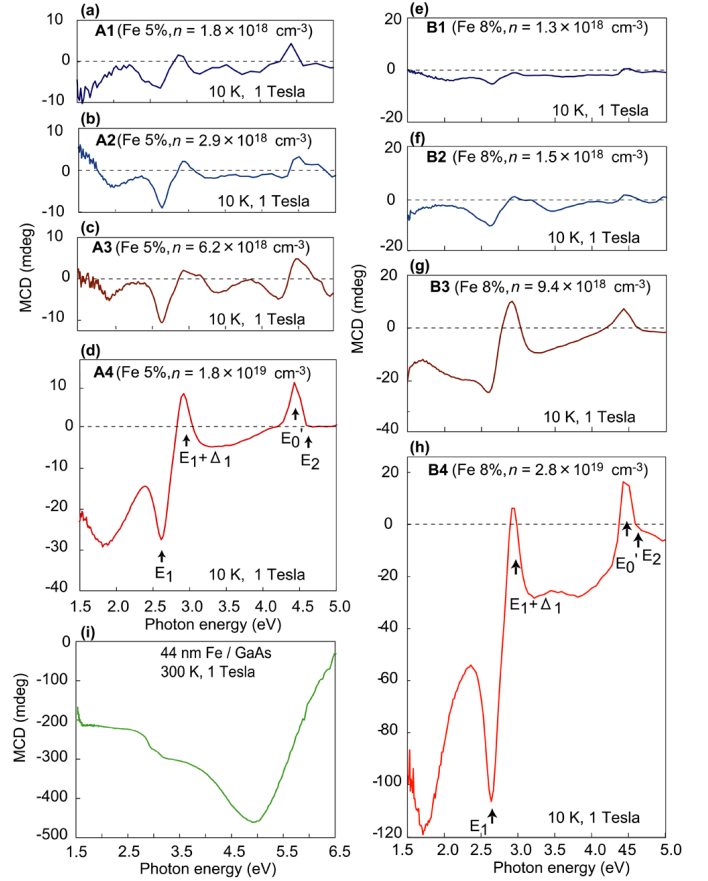

**FIG. 17.** (Color online) Magnetic circular dichroism (MCD) spectra of **(a)-(d)** (In$_{0.95}$,Fe$_{0.05}$)As samples (A1 - A4 in table I) with electron concentrations of $1.8 \times 10^{18}$, $2.9 \times 10^{18}$, $6.2 \times 10^{18}$, $1.8 \times 10^{19}$ cm$^{-3}$, respectively, measured at 10 K and under a magnetic field of 1 Tesla applied perpendicular to the film plane, and **(e)-(h)** (In$_{0.92}$,Fe$_{0.08}$)As samples (B1 - B4 in table I) with electron concentrations of $1.3 \times 10^{18}$, $1.5 \times 10^{18}$, $9.4 \times 10^{18}$, $2.8 \times 10^{19}$ cm$^{-3}$, respectively. With increasing the electron and Fe concentrations, the MCD spectra show strong enhancement at optical critical point energies $E_1$ (2.61 eV), $E_1 + \Delta_1$ (2.88 eV), $E_0$' (4.39 eV) and $E_2$ (4.74 eV) of InAs. **(i)** MCD spectrum of a 44 nm-thick Fe thin film grown on a GaAs substrate at 30°C, as a reference. The spectrum is clearly different from that of (In,Fe)As. Reprinted with permission from Appl. Phys. Lett. **101**, 182403 (2012). Copyright 2012 American Institute of Physics.

by the stray field of some embedded ferromagnetic Fe nanoclusters, if any. Furthermore, we show in Fig. 17 (i) the MCD spectrum of a 44 nm-thick epitaxial Fe thin film grown on a GaAs substrate at 30°C as a reference sample. The MCD spectra of our (In,Fe)As samples are clearly different from that of Fe, thus eliminating the possibility of metallic Fe particles. These results indicate that (In,Fe)As maintains its zinc-blende structure, and that its spin-split band structure is governed by the

*s,p-d* exchange interaction between the electron sea and the Fe magnetic moments. Samples A4, B3 and B4, whose electron concentrations are about $10^{19}$ cm$^{-3}$, are ferromagnetic, while other samples with lower electron concentrations are paramagnetic. These facts also indicate that the ferromagnetism of (In,Fe)As is induced by electrons and that we can rule out embedded metallic Fe or intermetallic Fe-As compound nanoparticles (if any) as the source of the observed ferromagnetism. At temperatures lower than 236°C, there are three intermetallic Fe-As compounds in their binary phase diagram: FeAs$_2$, FeAs and Fe$_2$As.[132] However, none of them is ferromagnetic; FeAs$_2$ is diamagnetic,[133] while FeAs and Fe$_2$As are both anti-ferromagnetic with Neel temperature of 77 K and $\sim$ 353 K, respectively.[132] We observed in-plane anisotropy in the magnetoresistance of a 10-nm thick n-type (In$_{0.94}$,Fe$_{0.06}$)As layer, and revealed a two-fold anisotropy along the [-110] direction and an 8-fold symmetric anisotropy along the crystal axes of (In,Fe)As, thus further supporting the intrinsic ferromagnetism of this material.[134]

In the following, we concentrate on the ferromagnetic behavior of sample A4 and B4. Figures 18 (a) and (b) show the normalized MCD spectra of sample A4 and B4, measured at 0.2, 0.5 and 1 Tesla. In Fig. 18 (a), the normalized MCD spectra of sample A4 show nearly perfect overlapping on a single spectrum over the whole photon-energy range, indicating that the MCD spectra comes from a single phase ferromagnetism of the whole (In,Fe)As film. In Fig. 18 (b), the normalized spectra of sample B4 shows perfect overlapping in the range of 2.5 – 5 eV, but deviate slightly from a single spectrum at photon energies lower than 2.5 eV. The peak at 1.8 eV develops faster than that at $E_1$ at low magnetic field, but they approach each other at 1 Tesla. The different behavior between sample A4 and B4 can be more clearly seen by plotting the normalized MCD intensity as a function of magnetic field (MCD-$H$ curve) at different photon energies (4.5 eV, 2.6 eV and 1.8 eV), as shown in Figs. 18 (d) and (e), respectively. While the magnetic field dependence of MCD of sample A4 measured at different photon energies perfectly agrees with each other (Fig. 18 (d)), that of sample B4 shows two different behaviors (Fig. 18 (e)). For sample B4, the MCD intensity at 1.8 eV reaches its saturation value at lower magnetic field than that at 2.6 and 4.5 eV. This shows that the MCD spectra of sample B4 come from two ferromagnetic phases. Both phases are zinc-blende-type (In,Fe)As with different Fe concentrations; one is the (In,Fe)As matrix phase having a MCD spectrum similar to that of sample A4 (Fig. 18 (a)), and the other is the cluster phase whose spectrum is enhanced at low photon energy (< 2.0 eV) as shown in Fig. 18 (c). The latter turned out to be zinc-blende (In,Fe)As clusters with higher density of Fe, as will be clarified later by SQUID measurements.

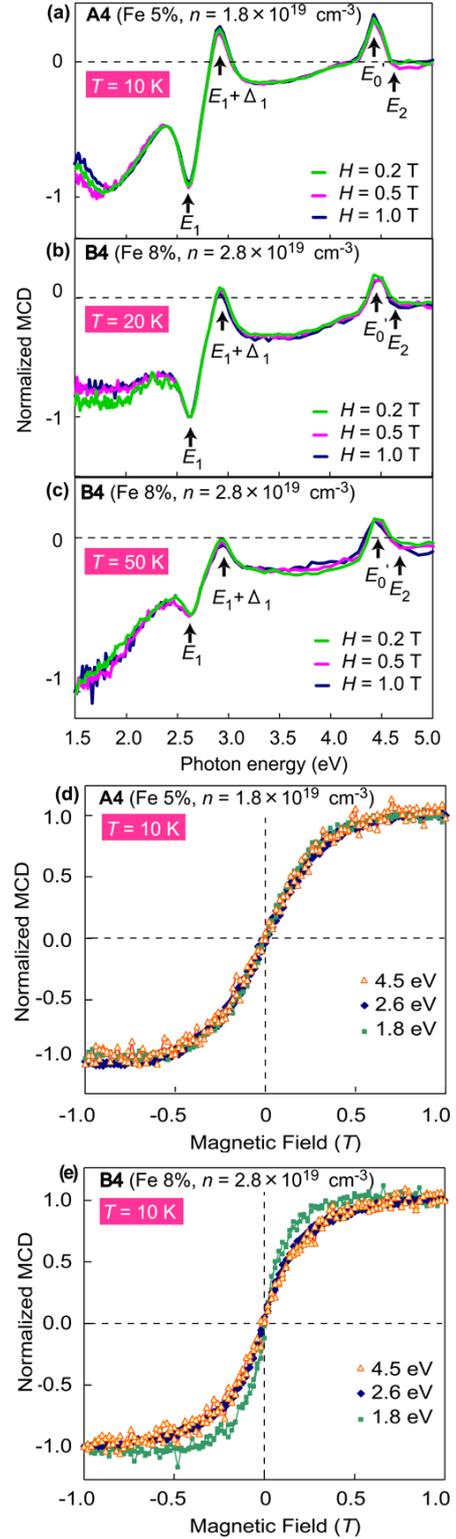

FIG. 18. (Color) **(a)** Normalized MCD spectra of sample A4 with $H$ = 0.2, 0.5 and 1 Tesla, measured at 10 K. A single spectrum over the whole photon-energy range indicates the single phase ferromagnetism of this sample. **(b)(c)** Normalized MCD spectra of sample B4 with $H$ = 0.2, 0.5 and 1 Tesla, measured at 20 and 50 K, respectively. The spectra at 20 K shows two phase ferromagnetism at low photon energy (< 2.0 eV), and can be decomposed to two components. One is the matrix with spectrum similar to the sample A4 as shown in (a), and the other is the cluster phase whose spectrum is enhanced at low photon energy (< 2.0 eV) as shown in (c). **(d)(e)** Normalized MCD – magnetic field (MCD - $H$) curves of samples A4 and B4, respectively, measured at photon energies of 1.8, 2.6 and 4.5 eV. The MCD - $H$ curves of sample A4 perfectly coincide with each other, while that of sample B4 at 1.8 eV shows smaller saturation field than that at 2.6 and 4.5 eV. Reprinted with permission from Appl. Phys. Lett. **101**, 182403 (2012). Copyright 2012 American Institute of Physics.

## E. Magnetization and magneo-optical imaging

The Arrott plots (MCD$^2$ vs. $H$/MCD) of sample A4 and B4 at different temperatures clearly show that sample A4 is ferromagnetic at $T < T_C \sim$ 34 K, and the matrix of sample B4 is ferromagnetic at $T < T_{C-1} \sim$ 28 K.[120] Next, we confirm these results by superconducting quantum interference device (SQUID) measurements. Figures 19 (a) and (b) show the field cooling (FC) and zero-field cooling (ZFC) magnetization-temperature ($M$-$T$) data of sample A4 and B4, measured by SQUID. The magnetic

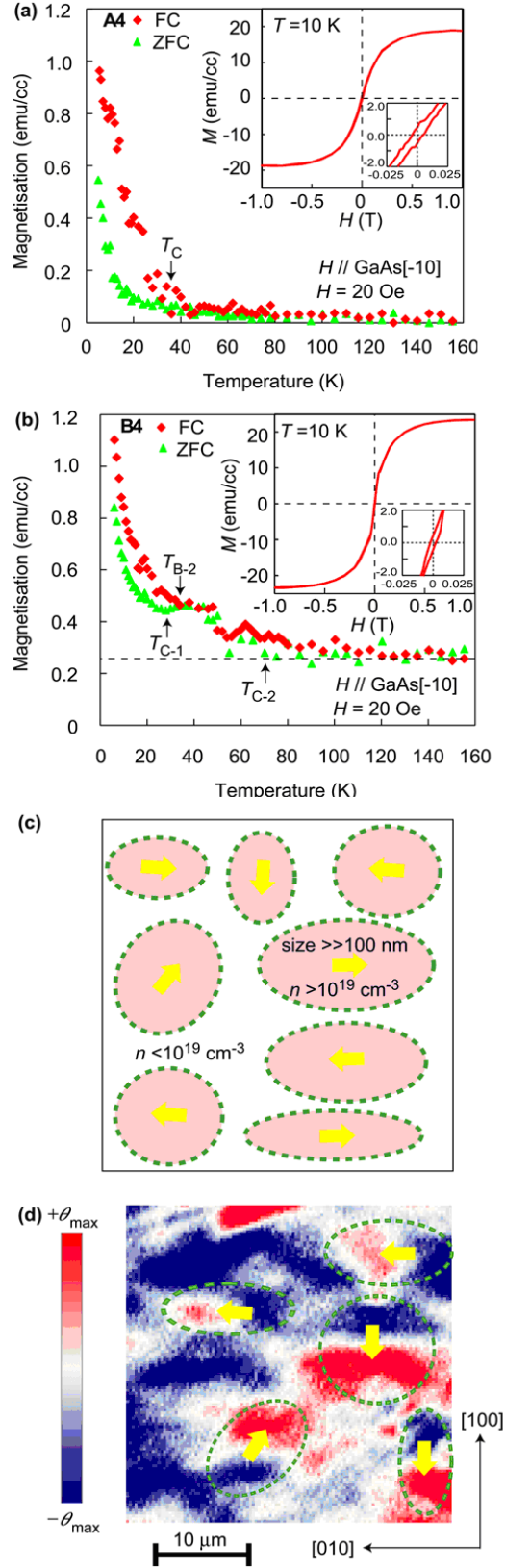

**FIG. 19.** (Color) **(a)(b)** Temperature dependence of the magnetization ($M$ - $T$ curves) of sample A4 and B4 measured by SQUID. The samples were cooled from room temperature to 5 K under two conditions, with a magnetic field of 1-Tesla (field-cooling, FC) and zero magnetic field (zero-field-cooling, ZFC). After cooling, the magnetization was measured with increasing temperature with a weak

magnetic field of 20 Oe. The magnetic field was applied in-plane along the GaAs[-110] direction. Sample A4 shows single-phase ferromagnetism with Curie temperature $T_C \sim 34$ K. In contrast, sample B4 shows two-phase ferromagnetism: One is the matrix phase with $T_{C-1} \sim 28$ K, and the other is the cluster phase with blocking temperature $T_{B-2} \sim 35$ K and Curie temperature $T_{C-2} \sim 70 \pm 10$ K. The insets show the magnetization hysteresis loops ($M$ - $H$) of sample A4 and B4 measured at 10 K. The magnified $M$ - $H$ curves near the origin are shown in the bottom-right of the insets, which clearly show the remnant magnetization and coercive force. **(c)** Discrete multi-domain model (plan view of the (In,Fe)As film). Pink areas (indicated by dashed lines) are ferromagnetic with $n > 10^{19}$ cm$^{-3}$, while white areas with $n < 10^{19}$ cm$^{-3}$ are not. Magnetization directions of ferromagnetic domains are indicated by yellow arrows. Each domain has a size much larger than the film thickness of 100 nm (see text). **(d)** Magneto-optical imaging of sample B4 under zero magnetic field at 4 K. The light source is a halogen lamp with a white light. Discrete ferromagnetic domains (shown by green dotted circles) with sizes of $\sim 10$ μm are visible. Areas with small Faraday rotation (white) between ferromagnetic domains correspond to paramagnetic areas. Reprinted with permission from Appl. Phys. Lett. **101**, 182403 (2012). Copyright 2012 American Institute of Physics.

field is applied in-plane along the GaAs[-110] direction. The $M$-$T$ curves of sample A4 show monotonous behavior both for FC and ZFC, which both rise at $T_C \sim 34$ K, indicating single-phase ferromagnetism. In contrast, sample B4 shows two-phase ferromagnetism; one is the matrix phase with $T_{C-1} \sim 30$ K, and the other is the superparamagnetic phase with $T_{B-2} \sim 35$ K and $T_{C2} \sim 70 \pm 10$ K. Note that the normalized MCD spectrum of sample B4 measured at 50 K (higher than $T_{C-1}$ and $T_{B-2}$ but lower than $T_{C2}$) still preserves clear features of the zinc-blende InAs structure (Fig. 18 (c)). This fact indicates that these magnetic clusters are *not* intermetallic precipitated particles but zinc-blende (In,Fe)As clusters with higher concentrations of Fe atoms. This is also consistent with the results of microstructure analysis shown in Fig. 15 (d), in which only zinc-blende structure is observed. The formation of the zinc-blende clusters with high concentration of magnetic atoms is the well-known spinodal decomposition phenomena,[135,136] which are observed in many FMSs such as (Ga,Mn)As[137,138], ZnCrTe[139] or GeFe[140] with high concentration of magnetic atoms. The $M$-$H$ curves measured with a magnetic field applied along the [-110] direction in the film plane are shown in the inset of Figs. 19 (a) and (b) for sample A4 and B4, respectively. The magnified in-plane $M - H$ curves near the origins (bottom-right of the insets of Figs. 19 (a) and (b)) clearly show hysteresis and remanent magnetization.

The remanent magnetizations of both $M$-$H$ curves (Figs. 19 (a) and (b)) are small. Here, we propose a discrete multi-domain model to explain such behavior of $M$-$H$ curves, as shown in Fig. 19 (c). In our model, the (In,Fe)As layer contains separate macroscopic ferromagnetic domains with sizes much larger than the film thickness of 100 nm. These ferromagnetic domains are separated from each other by paramagnetic areas in-between, and have different orientations of magnetization at zero magnetic field, resulting in small total remanent magnetization. The existence of paramagnetic areas between these domains are suggested by the fact that the average effective magnetic moments at saturation ($\mu_{eff} = 2.2$ and 1.7 $\mu_B$ per doped Fe atom for sample A4 and B4, respectively) are smaller than the expected $\mu_{Fe} = 5$ $\mu_B$ for Fe$^{3+}$, where $\mu_B$ is the Bohr magneton. Here, we used a magneto-optical (MO) imaging technique to visualize such ferromagnetic domains. Fig. 19 (d) shows a MO image of a local area (size 36 μm×36 μm) of sample B4, captured using an indicator garnet film in close contact with the surface of sample B4. The colored contrast in this image reflects the Faraday rotation of light going through the magnetic domains of the indicator garnet film whose magnetization is oriented by the stray fields from the underlying (In,Fe)As layer.[141] Many magnetic dipoles (shown by yellow arrows) with different magnetization orientations were observed, revealing the underlying ferromagnetic domains of (In,Fe)As with sizes of ~ 10 μm. There are also areas with nearly zero Faraday rotations (white areas) between these ferromagnetic domains, which correspond to the paramagnetic areas. The formation of these discrete multi-domain structures can be explained by the electron-induced ferromagnetism of (In,Fe)As and the phase separation due to the different concentrations of Be double donors and electron concentrations; areas with $n > 10^{19}$ cm$^{-3}$ (pink areas in Fig. 19 (c)) are ferromagnetic but areas with $n < 10^{19}$ cm$^{-3}$ (white) are not. The discrete ferromagnetic domains in electron-induced FMS is analogous to the discrete superconductive multi-domains observed in Cu-based high temperature superconductors. The calculated shape magnetic anisotropy ($1.0 \times 10^3$ kJ/m$^3$) of sample B4 based on our discrete multi-domain model is consistent with the measured value ($1.2 \times 10^3$ kJ/m$^3$).[120]

## F. Hall effect and magnetoresistance

In transport measurements, the electric currents in (In,Fe)As preferably flow through the ferromagnetic domains which have higher electron concentration and electrical conductivity. This allows us to examine the spin-dependent transport characteristics of the ferromagnetic domains in (In,Fe)As. The Hall effects, normal Hall effect and anomalous Hall effect (AHE), were measured in the Van der Pauw configuration. To eliminate the effect of the magnetoresistance due to misalignment of the Hall voltage terminals, we took the odd function from the raw data. To extract the normal and anomalous Hall effect components, we subtract the linear component (normal Hall effect $\sim \dfrac{1}{ne}$, where $n$ is the electron concentration) from the raw Hall effect data, so that the remaining non-linear component (AHE $\sim M$, where $M$ is the magnetization) has the same zero-field susceptibility as that obtained by SQUID magnetometry or MCD. Figs. 20 (a) and (b) show the Hall resistance of sample A4 and B4, respectively. The normal Hall effect with negative gradient, indicating the n-type conduction of these (In,Fe)As layers, dominates the Hall resistance. The n-type conductivity is also confirmed by the polarity of the thermoelectric Seebeck coefficient (see section H). There is a small fraction (~ 3%) of positive AHE contribution in both samples due to the spin-dependent scattering of electrons at Fe sites, as shown in the anomalous Hall resistance (AHR) curves in Figs. 20 (c) and (d). The weak AHE in n-type FMS compared with that of p-type FMS is consistent with the Berry-phase theory of AHE in FMS.[142] The normalized AHE curve of sample A4 perfectly agrees with those of MCD and magnetization as shown in Fig. 20 (e), indicating again that the ferromagnetism in this sample comes only from the single phase. In contrast, the results of sample B4 are more complicated. In Fig. 20 (f), the normalized $M$-$H$ curve measured by SQUID lies in the middle of the MCD-$H$ curve measured at 1.8 eV (dominated by the superparamagnetic cluster phase) and 2.6 eV (dominated by the matrix phase). This is reasonable since SQUID measures averaged signals from all phases in the sample, while MCD can selectively pick up different signals from different phases by changing the photon energy. This fact demonstrates the advantage of the MCD technique in our study. The normalized AHE curve of sample B4 agrees well with the normalized MCD at 2.6 eV at magnetic field smaller than 0.3 Tesla, suggesting that the spin-dependent scattering in the matrix mainly contributes to the AHE at low magnetic field. At magnetic field higher than 0.3 Tesla, the normalized AHE is deviated from the normalized MCD at 2.6 eV, which can be attributed to its two-phase structure.

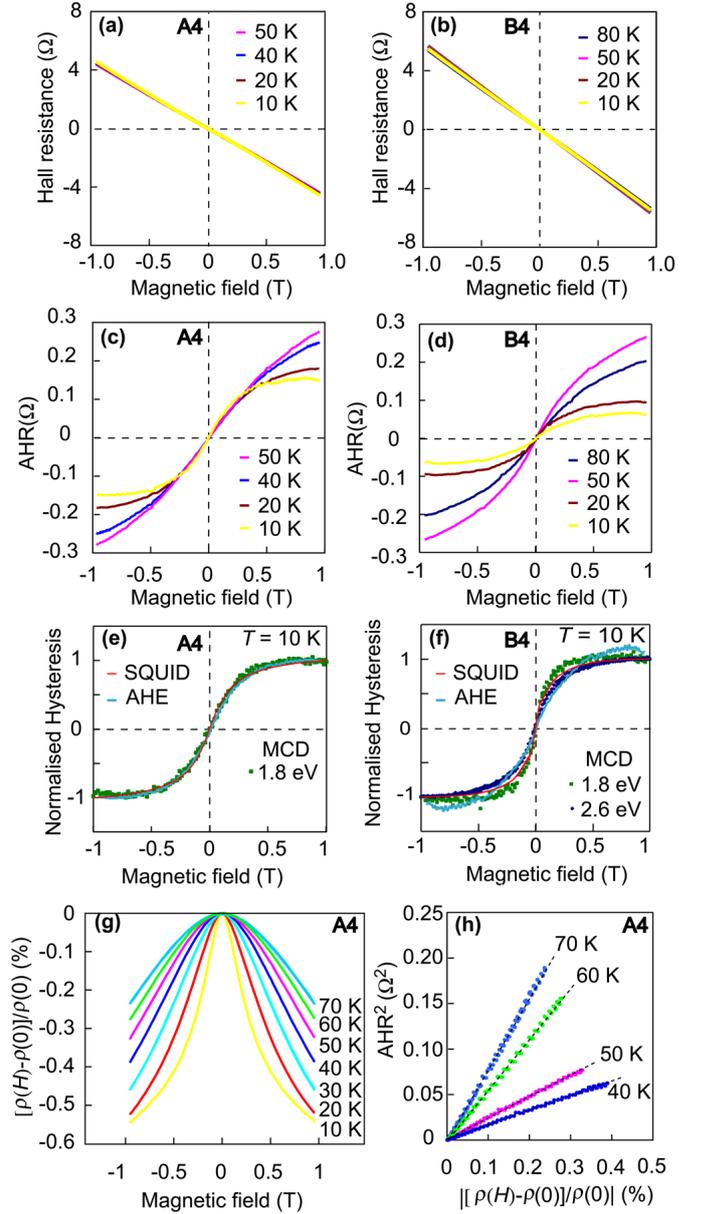

FIG. 20. (Color) **(a)(b)** Hall resistances of sample A4 and B4 at various temperatures, respectively. The Hall resistance of (In,Fe)As is dominated by the normal Hall effect with negative gradient, showing the n-type conduction of these samples. **(c)(d)** Extracted *positive* anomalous Hall resistances (AHR) for sample A4 and B4, respectively. The AHR are about 3% of the normal Hall resistances, even at 10 K. Nevertheless, clear temperature dependence of these AHR was observed. **(e)(f)** Comparison of the magnetic field dependence of MCD, magnetization, and AHR at 10 K for sample A4 and B4, respectively. **(g)** Magnetoresistance of sample A4, normalized by its value at zero magnetic field. The magnetic field is applied perpendicular to the film plane. **(h)** Magnetoresistance ratio $|[\rho(H) - \rho(0)]/\rho(0)|$ vs. AHR$^2$ of sample A4 measured at $40 - 70$ K. Excellent linear relationships indicate that both AHE and magnetoresistance originate from the spin-dependent scattering in the (In,Fe)As layer. (a) – (f) are reprinted with permission from Appl. Phys. Lett. **101**, 182403 (2012). Copyright 2012 American Institute of Physics.

Furthermore, we examined the magnetoresistance in sample A4 to find further evidence of spin-dependent scattering. Fig. 20 (g) shows the magnetoresistance $|[\rho(H) - \rho(0)]/\rho(0)|$ of sample A4 measured at various temperatures, where $\rho(H)$ and $\rho(0)$ are the resistivity at a magnetic field of $H$ and 0, respectively. Clear negative magnetoresistance was observed in the whole temperature range ($T$ = 10 - 70 K). The negative magnetoresistance can be understood as the reduction of spin-disorder scattering when the magnetic moments of Fe atoms are aligned along $H$. Above the Curie temperature (34 K), where the spin-spin correlation between Fe atoms are weak, the magnetoresistance ratio $|[\rho(H) - \rho(0)]/\rho(0)|$ is proportional to $M^2$. Since AHR is proportional to $M$ as evidenced in Fig. 20 (e), a linear relationship between $|[\rho(H) - \rho(0)]/\rho(0)|$ and $AHR^2$ should be expected. Fig. 20 (h) shows $|[\rho(H) - \rho(0)]/\rho(0)|$ vs. $AHR^2$ plotted at $T$ = 40 – 70 K. Excellent linear relationships between $|[\rho(H) - \rho(0)]/\rho(0)|$ and $AHR^2$ are observed, indicating that both the observed AHE and negative magnetoresistance originate from the spin-dependent scattering in this (In,Fe)As sample.

## G. Electron-induced ferromagnetism in (In,Fe)As

In Figs. 21 (a) and (b), we show the evolution of ferromagnetism expressed by $T_C$ vs. electron concentration $n$ and resistivity vs. temperature of sample series A and B, respectively. It is clear that there is a threshold electron concentration of about $10^{19}$ cm$^{-3}$ for (In,Fe)As to become ferromagnetic. The steep change in magnetic behavior at $10^{19}$ cm$^{-3}$ shown in Fig. 21 (a) is clearly correlated with the metal-insulator transition of (In,Fe)As layers as shown in Fig. 21 (b). All of these results confirm that (In,Fe)As is an intrinsic n-type ferromagnetic semiconductor whose ferromagnetism is induced by electrons. It should be noted that sample A4 with $T_C$ = 40 K requires an electron concentration of $1.8 \times 10^{19}$ cm$^{-3}$. Comparing with (In,Mn)As, this electron concentration is an order of magnitude smaller ($T_C$ ~ 20 K requires 1.0 - $1.6 \times 10^{20}$ cm$^{-3}$ of holes for (In,Mn)As, see Ref.[115]). Noting that a carrier concentration change of ~ $10^{20}$ cm$^{-3}$ can be obtained by applying a gate voltage in field-effect transistor structures,[59] this smaller electron concentration gives (In,Fe)As another advantage over (In,Mn)As when controlling ferromagnetism by electrical and optical means.

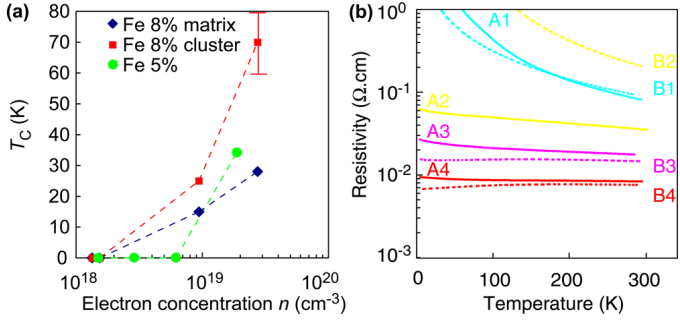

**FIG. 21.** (Color) **(a)** $T_C$ vs. electron concentration and **(b)** resistivity vs. temperature summarized for sample series A and B. An electron concentration threshold of about $1 \times 10^{19}$ cm$^{-3}$ is needed for ferromagnetism, which is also the boundary for metal-insulator transition. Reprinted with permission from Appl. Phys. Lett. **101**, 182403 (2012). Copyright 2012 American Institute of Physics.

What can be expected using an n-type FMS? There have been already a large number of proposed spin-devices using pn junctions with a p-type FMS and non-magnetic n-type semiconductor or vice-versa, in which carrier spins in non-magnetic layers are generated by irradiating circularly polarized light.[143 -146] With an n-type FMS, we can realize many spin-devices such as spin diodes,[143] spin bipolar transistors,[146,147] and spin metal-oxide- semiconductor field-effect transistors (spin MOSFETs),[10,148] without using any external light source. Those devices can be used for high-density non-volatile magnetic memory and reconfigurable logic circuits.[11,149]

Using Fe as magnetic dopants has another important advantage over Mn, especially when studying the mechanism of carrier-induced ferromagnetism. In the case of (Ga,Mn)As, there are Mn-related impurity states, which complicate the theory of carrier-induced ferromagnetism. In contrast, Fe atoms in III-V are neither major donors nor acceptors; thus, there are probably no available or active Fe-related donor or acceptor impurity states. The original mean-field Zener model of carrier-induced ferromagnetism in Mn based FMSs was developed based on the assumption that holes reside in the valence band (VB).[47, 56] This model has been widely used as the standard theory of carried-induced ferromagnetism in Mn based FMS, since it can explain some features of (Ga,Mn)As.[58,150] On the other hand, recent reports on the optical[61,62,63] and transport properties of (Ga,Mn)As have shown that holes exist in the impurity band within the band gap of (Ga,Mn)As with an effective mass as heavy as $10m_0$, where $m_0$ is the free electron mass. Those results make the assumption of the mean-field Zener model unjustified, and suggest an alternative model called the impurity band (IB) model, as discussed in the previous chapter. The difficulty in understanding the ferromagnetism in (Ga,Mn)As comes from the existence of such IB in the band gap, with which it is difficult to deal theoretically and experimentally. In the next section, we show that electrons in (In,Fe)As are in the conduction band of InAs, and not related to any Fe hypothetical d-band or itinerant impurity states. This may greatly reduce the complexity of interpretation of the ferromagnetism in this material.

## H.   Thermoelectric Seebeck effect, electron effective mass, and Fermi level[121]

A convenient way to confirm the carrier type and estimate the effective mass in heavily doped semiconductors is the thermoelectric Seebeck effect measurement. When there is a temperature gradient $\Delta T$ between two edges of a sample, carriers at the hot side are more thermally activated and then diffuse to the cold side, resulting in a voltage $\Delta V$ that will be generated between the two edges. Equilibrium is established when this voltage is sufficient to stop further net carrier diffusion. The Seebeck coefficient $\alpha$ of a material is defined as $\alpha = -\Delta V / \Delta T$. If carriers are electrons (holes), $\alpha$ is negative (positive). Fig. 22 (a) shows the experimental setup to measure the Seebeck effect of our (In,Fe)As at room temperature. The hot side is a copper (Cu) electrode with a heater, placed on an epoxy film. The epoxy film acts as a thermal insulator. The cold side is a Cu electrode placed on a sapphire substrate, which acts as a thermal sink. A piece of sample bridges the hot and cold electrodes. Silver paste is used for electrical contacts between the edges of the sample and the electrodes. Voltage signals from a thermocouple made from a Cu wire (red line) and Constantan wire (blue line) measure the temperature difference $\Delta T_{raw}$ between the hot Cu electrode and the sapphire substrate when the heater is turned on. Fig. 22 (b) shows the measured $\Delta V$ - $\Delta T_{raw}$ of sample B4 at different heater currents. It is clear that $\alpha$ is negative from the gradient of this data. Thus, the majority carriers are electrons, which is consistent with the Hall effect measurement results described in section F.

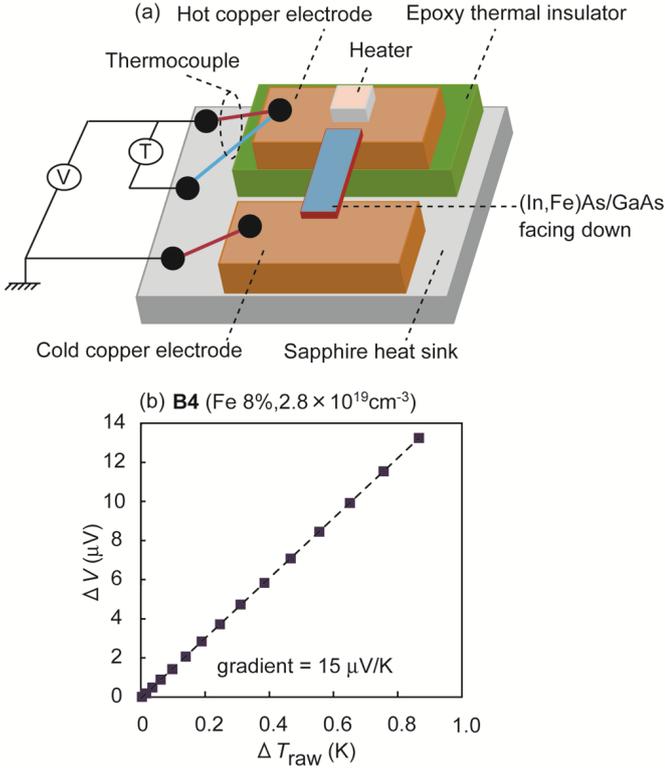

**FIG. 22.** (Color) **(a)** Experimental setup for the Seebeck effect measurements of the (In,Fe)As thin films grown on semi-insulating GaAs substrates. All the measurements are carried out at room temperature. **(b)** Measured $\Delta V$ - $\Delta T_{raw}$ of sample B4. Reprinted with permission from Appl. Phys. Lett. 101, 252410 (2012). Copyright 2012 American Institute of Physics.

Using the values of $\alpha$ and electron concentration $n$, we can estimate the effective mass $m^*$ of electrons and the Fermi energy $E_F$ (with respect to the conduction band bottom) by using the following equations:

$$\alpha = -\frac{k_B}{e}\frac{\pi^2}{3}\left(s + \frac{3}{2}\right)\frac{k_B T}{E_F}, \qquad (1)$$

$$n = \frac{4N_C}{3\sqrt{\pi}}\left(\frac{E_F}{k_B T}\right)^{3/2}, \qquad (2)$$

$$N_C = 2\left(\frac{m^* k_B T}{2\pi\hbar^2}\right)^{3/2}. \qquad (3)$$

Here $k_B$ is the Boltzmann constant, $e$ is the elementary charge, $N_C$ is the effective density of state. $s$ in Eq. (1) is the exponent of the scattering time $\tau \sim \varepsilon^s$. Here we use $s = 0$ for neutral impurity scattering.

The electron concentration $n$ can be easily obtained from the Hall effect measurement at room temperature. Note that the anomalous Hall effect is quite small compared with the normal Hall effect even at low temperature, so we can neglect its contribution at room temperature in this material. The magnitude of $\alpha$ is given by

$$\alpha = -\frac{\Delta V}{\Delta T_{edge}} = -\left(\frac{\Delta T_{raw}}{\Delta T_{edge}}\right)\frac{\Delta V}{\Delta T_{raw}} = -k\frac{\Delta V}{\Delta T_{raw}},$$

where $k$ is the ratio between the measured $\Delta T_{raw}$ and the real temperature difference $\Delta T_{edge}$ between the two edges of the sample. If the thermal conductivity of a sample is much smaller than those of copper and sapphire, then $\Delta T_{raw} = \Delta T_{edge}$. In reality, due to the good thermal conductivity of GaAs, there is a temperature distribution in the electrodes and sapphire substrate. As a result, $\Delta T_{edge}$ is generally smaller than $\Delta T_{raw}$. $k$ is measured to be 2 for a reference sapphire sample, whose thermal conductivity ~ 0.42 W/(cm·degree)

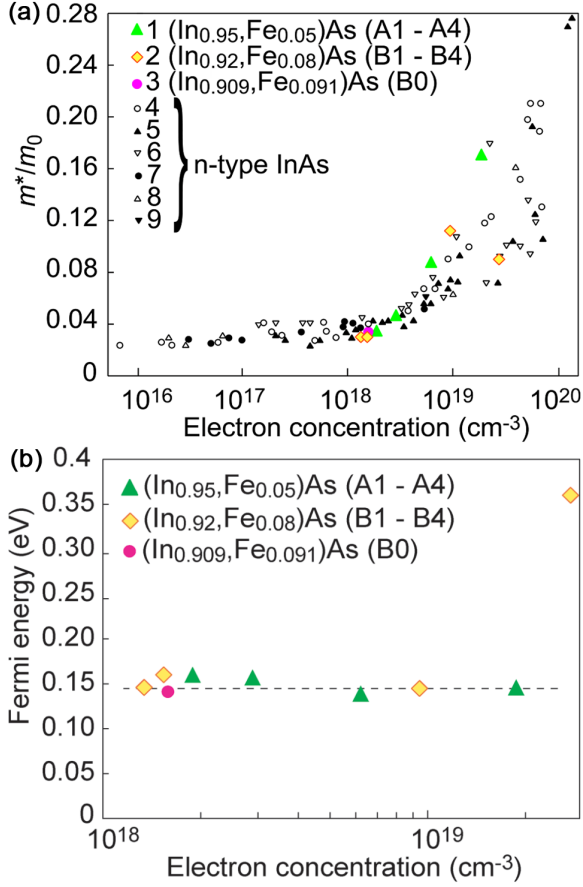

(a) Graph: $m^*/m_0$ vs Electron concentration (cm$^{-3}$)

1 (In$_{0.95}$,Fe$_{0.05}$)As (A1 - A4)
2 (In$_{0.92}$,Fe$_{0.08}$)As (B1 - B4)
3 (In$_{0.909}$,Fe$_{0.091}$)As (B0)
4
5
6
7
8
9

n-type InAs

(b) Graph: Fermi energy (eV) vs Electron concentration (cm$^{-3}$)

(In$_{0.95}$,Fe$_{0.05}$)As (A1 - A4)
(In$_{0.92}$,Fe$_{0.08}$)As (B1 - B4)
(In$_{0.909}$,Fe$_{0.091}$)As (B0)

**FIG. 23.** (Color)   **(a)** Electron effective mass $m^*$ vs. electron concentration $n$ for various InAs-based semiconductors. Our data (colored symbols) are obtained by the Seebeck effect for (1) Be doped (In$_{0.95}$,Fe$_{0.05}$)As samples (A1 - A4), (2) Be doped (In$_{0.92}$,Fe$_{0.08}$)As samples (B1 - B4), and (3) undoped (In$_{0.909}$,Fe$_{0.091}$)As (sample B0). Black and white symbols are $m^*$ of n-type InAs obtained by (4) the Seebeck effect, (5) infrared reflectivity, (6) magnetic susceptibility, (7) Faraday effect, (8) recombination radiation, and (9) cyclotron resonance (after Ref. 151 and references therein).   **(b)** Fermi energy $E_F$ of (In,Fe)As as a function of electron concentration $n$. The dashed line indicates that $E_F$ increases very slowly due to the rapid increase of $m^*$ with $n$. Reprinted with permission from Appl. Phys. Lett. 101, 252410 (2012). Copyright 2012 American Institute of Physics.

is nearly equal to 0.44 W/(cm·degree) of semi-insulating GaAs. Therefore, in this experiment, we multiply the gradient of $\Delta V$ - $\Delta T_{\text{raw}}$ data by $-k = -2$ to obtain the magnitude of $\alpha$. For example, $\alpha$ of sample B4 is estimated to be -30 μV/K from the data of Fig. 22 (b).

Color plots in Fig. 23 (a) show the obtained effective mass $m^*$ of our several (In,Fe)As samples (series A and B in this work)

with varying the Fe concentration and electron concentration. It is found that $m^*$ is 0.030 ~ 0.171$m_0$ depending on the electron concentration. These $m^*$ values are all close to the effective mass values of the conduction band electrons reported in heavily doped InAs (black and white dots in Fig. 23 (a)),[151] indicating that the electrons in (In,Fe)As reside in the conduction band, not in the hypothetical Fe-related impurity band with heavy effective mass. This result is a stark contrast to that of holes in (Ga,Mn)As, whose effective mass is found to be of the order of 10$m_0$ by infrared absorption spectroscopy.[61] Sample B4 has relatively small $m^*$ despite large $n$, which is probably due to the local fluctuation of $n$ in this sample.

Fig. 23 (b) shows $E_F$ of all the samples as a function of $n$ derived by Eq. (1). We found that the $E_F$ values are at least 0.15 eV. Except for sample B4, $E_F$ is almost unchanged for the samples with $n$ between 1.3×10$^{18}$ cm$^{-3}$ and 1.8×10$^{19}$ cm$^{-3}$, reflecting the experimental fact that $\alpha$ is almost unchanged (~ −70 μV/K). In theory, $E_F$ should increase with increasing $n$. However, the data in Fig. 23 (b) show that the increase of $E_F$ is very little. This can be understood as follows: Since $m^*$ quickly increases from 0.03$m_0$ to 0.17$m_0$ when $n$ increases from 1.3×10$^{18}$ cm$^{-3}$ to 1.8×10$^{19}$ cm$^{-3}$, the density of states $\rho = m^* k_F / \pi^2 \hbar^2$ at the Fermi level with the Fermi wavenumber $k_F$ rapidly increases with increasing $n$; thus, the increase of $E_F$ itself is compensated. Sample B4, which has the highest $n$, has an apparently large $E_F$ of 0.36 eV due to its small $m^*$ (the reason for small $m^*$ and large $E_F$ in sample B4 is described in Supplementary Material in Ref. 121). The results of Fig. 23 (b) indicate that $E_F$ in our (In,Fe)As samples is quite large (0.15 − 0.36 eV). Since the band gap of InAs is only 0.36 eV at room temperature, if the electron gas resided in a hypothetical narrow $d$-like impurity band inside the band gap, $E_F$ would not be that large. Therefore, we again conclude that electron carriers in (In,Fe)As reside in the conduction band. Our conclusion is consistent with the fact that the non-crystalline anisotropy magnetoresistance (AMR) ratio in (In,Fe)As is as small as 0.001%. [134] Because the non-crystalline AMR reflects the scattering processes of electrons from $s$ states to $d$ states at the Fermi level, this very small non-crystalline AMR supports our "conduction band" scenario, in which there are only scattering processes from $s$ states to other $s$ states. Very recently, this "conduction band" scenario in (In,Fe)As has been directly confirmed by angle-resolved photoemission spectroscopy, in which $E_F \sim 0.15$ eV lies in the conduction band and this $E_F$ value is nearly the same in samples with small $n \sim 1 \times 10^{18}$ cm$^{-3}$ and large $n \sim 1 \times 10^{19}$ cm$^{-3}$, as estimated in Fig. 23 (b). [152]

## I. Large s-d exchange

The results described in the previous section indicates that (In,Fe)As is free from the "impurity band" problem that complicates the theoretical interpretations of ferromagnetism in Mn-based FMSs. It is therefore reasonable to apply the Zener model originally developed for "valence band" p-type FMSs to estimate the $s$-$d$ exchange interaction in (In,Fe)As. According to this model, $T_C$ is given by $T_C = xN_0 S(S+1)\alpha^2 A_F \rho /12k_B$. [47,56] Here, $S = 5/2$ is the Fe local spin, $A_F$ is the Fermi liquid parameter. Using the data of sample A4 ($x = 5\%$, $T_C = 34$ K, $n = 1.8 \times 10^{18}$ cm$^{-3}$, $m^* = 0.17m_0$) and assuming that $A_F = 1$, the exchange interaction $|N_0\alpha|$ is estimated to be 2.8 eV. This value is not only one order of magnitude larger than that ($\sim 0.2$ eV) of II-VI based DMSs, but also larger than that of the $p$-$d$ exchange interaction in (Ga,Mn)As ($|N_0\beta| \sim 1.2$ eV) [153]. This large value of $|N_0\alpha|$ reflects the fact that $T_C$ in (In,Fe)As can be as high as several tens of Kelvin when $n$ is as low as $10^{19}$ cm$^{-3}$, while $T_C$ of the same order in Mn-based FMSs often requires a much higher hole concentration of $\sim 10^{20}$ cm$^{-3}$.

We point out that the observed relatively high $T_C$ (tens of Kelvin at $n \sim 10^{19}$ cm$^{-3}$) and the large $s$-$d$ exchange interaction are not expected from the chemical trend extrapolated either from that of II-V based DMSs or from Mn-doped III-V based FMSs. [47,56] The observed ferromagnetism with relatively high $T_C$ at low $n$ in (In,Fe)As is therefore surprising. One may suspect the possibility of metallic Fe or inter-metallic Fe-As nanoparticles, but such a possibility has been ruled out by our careful structural characterizations. Furthermore, the intrinsic ferromagnetism of (In,Fe)As has been confirmed by various experiments including the anomalous Hall effect, MCD spectroscopy, magneto-optical imaging of ferromagnetic domains, crystalline AMR and planar Hall effect, and the striking dependence of $T_C$ on $n$ at fixed Fe concentrations. Those experimental results, as described in the previous sections, cannot be explained by second-phase precipitates.

Although the physical origin of this large $s$-$d$ exchange interaction in n-type (In,Fe)As is not clear at this stage, we can think of a possible scenario that may explain the large $s$-$d$ exchange interaction in n-type FMSs. The $s$-$d$ exchange interaction energy derived from the Anderson Hamiltonian [154] is given by [155]

$$N_0\alpha = -2 |V_{sd}|^2 \left( \frac{1}{E_C - \varepsilon_d} + \frac{1}{U - E_C + \varepsilon_d} \right).$$

Here, $E_C$ is the energy of the conduction band bottom, $\varepsilon_d$ is the energy of the $d$ states of Fe, $U$ is the Coulomb repulsion between opposite-spin electrons in a $d$ state, and $V_{sd}$ is the $s$-$d$ mixing potential. It is commonly said that $V_{sd}$, and thus $N_0\alpha$, should be small, because the $s$ wavefunctions of electrons are orthogonal to the localized $d$ wavefunctions of magnetic atoms. However, as pointed out by Anderson[154], while a $d$ orbital is orthogonal to a $s$ orbital of the same atom, it is generally not orthogonal to a $s$ orbital of the neighboring atoms. In the case of diluted magnetic alloys such as CuMn or AgMn, the $|V_{sd}|$ value was experimentally obtained in the range of 0.4 - 2.5 eV. [156] Furthermore, if the energy difference between the Fe $d$ states and the conduction band ($E_C - \varepsilon_d$) is very small (here referred to as "resonance"), as shown in Fig. 24, the term in the brackets could be remarkably large. We suggest the bandstructure of (In,Fe)As as shown in Fig. 24. In this picture, if we take $|V_{sd}| \sim 1$ eV, $E_C - \varepsilon_d \sim 0.5$ eV, and neglect the second part in the bracket (since $U$ is typically very large), we can obtain $N_0\alpha = -4$ eV. Such a resonant situation may occur in (In,Fe)As. In the resonant case, the empirical chemical trend of exchange interactions suggested in Refs. 47 and 56 would be no longer be applicable.

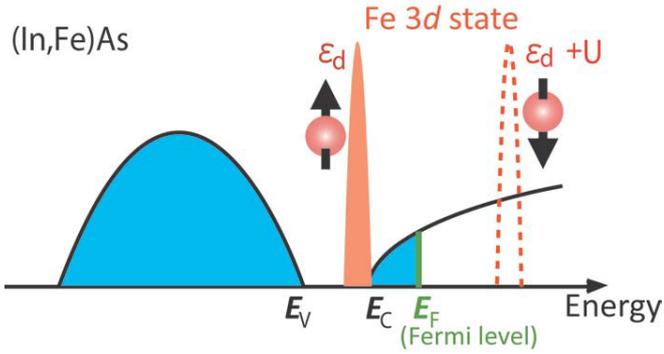

**FIG. 24.** (Color online) Schematic band structure of (In,Fe)As. The overlap of the $d$-level of the Fe impurities and the conduction band possibly induces the large $s$-$d$ exchange interaction in (In,Fe)As.

## IV. SUMMARY AND OUTLOOK

We described the recent progress in III-V based FMSs with focus on the tunneling transport in (GaMn)As based heterostructures, and resonant tunneling spectroscopy leading to the characterization of the bandstructure and the Fermi level position in (GaMn)As. First, we showed that large TMR was clearly observed in single-barrier MTJs with (GaMn)As electrodes. The tunnel barrier material was AlAs, GaAs, and AlMnAs. The basic properties of such all-semiconductor MTJs were presented, including the barrier thickness dependence of RA and TMR ratio, and temperature and bias dependence of TMR. This study was extended to the spin-dependent resonant tunneling in RTD structures containing a GaMnAs QW and double tunnel barriers. We observed clear resonant tunneling through the quantum levels in a GaMnAs QW sandwiched by double barriers. Enhancement of TMR was clearly observed at resonant levels. We carried out systematic experiments using resonant tunneling spectroscopy on double-barrier tunnel junctions with various GaMnAs QW thicknesses and performed theoretical analyses/fittings. It was found that the Fermi levels of (GaMn)As is located, not in the VB, but always in the bandgap in all the insulating and metallic (GaMn)As with Mn concentrations from 0.01% to 15%. Our study indicates that VB of (GaMn)As is nearly nonmagnetic with small spin splitting, not largely affected by Mn doping, and that IB plays an important role in transport and ferromagnetism. We also mentioned some device applications using GaMnAs-based hetero-structures; MTT, three-terminal RTD with a GaMnAs QW, and double-quantum well heterostructure. If we can better control the IB nature and increase $T_C$ in the future, these devices may be applied to magnetic-field sensors, nonvolatile memory/logic, and reconfigurable logic devices, which are compatible with III-V electronics.

Furthermore, we described the growth of n-type FMS, (InFe)As, and its structural, magnetic, transport, and magneto-optical properties. By introducing isoelectronic Fe magnetic impurities and Be double-donor atoms into InAs, we have grown n-type ferromagnetic (InFe)As thin films with the ability to control ferromagnetism by both Fe and independent carrier doping. It is shown that (In,Fe)As doped with electrons behaves as an n-type electron-induced FMS. MCD, SQUID, and magnetotransport data indicate clear evolution of ferromagnetism in (In,Fe)As when increasing the electron density by Be doping with a fixed Fe concentration. We have also estimated the electron effective mass and the Fermi energy of n-type (In,Fe)As with different electron concentrations. The obtained very small electron effective mass ($m^*$=0.03 - 0.17$m_0$) and large Fermi energy (at least 0.15 eV) indicate that electron carriers reside in the conduction band of (In,Fe)As, rather than in a hypothetical Fe-related itinerant IB. Using the experimental value of $m^*$ and the conventional theory, the s-d exchange interaction $|N_0\alpha|$ is estimated to be 2.8 eV, much larger than the values reported in conventional DMS materials. Although further theoretical and experimental investigations are needed, this means that (In,Fe)As is promising for high-$T_C$ FMS if the electron concentration can be increased up to ~$10^{20}$ cm$^{-3}$. Development of such Fe-based n-type electron-induced FMS, together with Mn-based p-type hole-induced FMS, will open the way to fabricate all-FMS spintronic devices, as well as helps to understanding the physics of carrier-induced ferromagnetism in semiconductors.

**Acknowledgements**: The authors wish to thank Iriya Muneta, Le Duc Anh, Kenta Takata, Hiroshi Terada, and Daisuke Sasaki in our group for their contributions to this work. Thanks are also due to Dr. S. Mohan and Prof. T. Tamegai of Dept. of Applied Physics at the University of Tokyo for magneto-optical imaging measurements, Dr. M. Kodzuka, Dr. T. Ohkubo, and Dr. K. Hono of NIMS for 3DAP measurements, and Dr. T. Fujii and staffs of Cryogenic Center at the University of Tokyo for the help in the Seebeck effect and SQUID measurements. This work was partly supported by Grant-in-Aids for Scientific Research, particularly Specially Promoted Research, Project for Developing Innovation Systems of MEXT, FIRST Program of JSPS, and Global COE/Outstanding Graduate School Programs. One of the authors (PNH) acknowledges the support from Foundation of Ando Laboratory.

## References

[1] See for example, *Special Issue on Spintronics*, IEEE Trans. Electron Dev. **54**, No. 5 (2007).

[2] I. Zutic, J. Fabian, and S. Das Sarma, Rev. Mod. Phys., **76**, 323 (2004).

[3] M.N. Baibich, J.M. Bruto, A. Fert, F. Nguyen, van Dau, F. Petroff, P. Eitenne, G. Creuzet, A. Friederich and J. Chazelas, Phys. Rev. Lett. **61**, 2472 (1988).

[4] G. Binasch, P. Grünberg, F. Saurenbach and W. Zinn, Phys. Rev. B **39**, 4828 (1989).

[5] M. Julliere, Phys. Lett., **54A**, 225 (1975)

[6] T. Miyazaki and N. Tezuka, J. Magn. Magn. Mat., **139**, L231 (1995).

[7] J. S. Moodera, L. R. Kinder, T. M. Wong, and R. Meservey, Phys. Rev. Lett., **74**, 3273 (1995).

[8] M. Tanaka and S. Ohya, *Spintronic Devices Based on Semiconductors* in P. Bhattacharya, R. Fornari, and H. Kamimura (eds.), Comprehensive Semiconductor Science and Technology, vol. **6**, pp. 540–562, Amsterdam: Elsevier, February 2011.

[9] S. Datta S and B. Das, Appl. Phys. Lett. **56**, 665 (1990).

[10] S. Sugahara and M. Tanaka, Appl. Phys. Lett. **84**, 2307 (2004).

[11] M. Tanaka and S. Sugahara, IEEE Trans. Electron Dev. **54**, 961 (2007).

[12] T. Kasuya and A. Yanase, Rev. Mod. Phys. **40**, 684 (1968).

[13] J. Furdyna, J. Appl. Phys. **64**, R29 (1988).

[14] T. Story, R.R. Galazka, R.B. Frankel, P.A. Wolff, Phys. Rev. Lett. **56**, 777 (1986).

[15] H. Munekata, H. Ohno, S. von Molnar, A. Segmuller, L. L. Chang and L. Esaki, Phys. Rev. Lett. **63**, 1849 (1989).

[16] H. Ohno, H. Munekata, S.von Molnar and L.L.Chang, J. Appl. Phys. **69**, 6104 (1991).

[17] H. Munekata, H. Ohno, R. R. Ruf, R. J. Gambino and L. L. Chang, J. Cryst. Growth **111**, 1011 (1991).

[18] H. Ohno, H. Munekata,T. Penny, S.von Molnar and L. L. Chang, Phys. Rev. Lett. **68**, 2664 (1992).

[19] H. Ohno, A. Shen, F. Matsukura, A. Oiwa, A.

Endo, S. Katsumoto and H. Iye , Appl. Phys. Lett. **69**, 363 (1996).

[20] T. Hayashi, M. Tanaka, T. Nishinaga, H. Shimada, H. Tsuchiya and Y.Ootuka, J. Cryst. Growth **175/176**, 1063 (1997).

[21] A.Van Esch, L.Van Bockstal, J. De Boeck, G.Verbanck, A. S. van Steenbergen, P. J. Wellmann, B. Grietens, R. B. F. Herlach and G. Borghs, Phys. Rev. B **56**, 13103 (1997).

[22] T. Jungwirth, J. Sinova, J. Macek, J. Kucera, and A. H. MacDonald, Rev. Mod. Phys. **78**, 809 (2006).

[23] N. Samarth, Nature Mat. **11**, 360 (2012).

[24] T. Hayashi, H. Shimada, H. Shimizu, and M. Tanaka, J. Cryst. Growth **201/202**, 689 (1999).

[25] D. Chiba, N. Akiba, F. Matsukura, Y. Ohno, and H. Ohno, Appl. Phys. Lett. **77**, 1873 (2000).

[26] M. Tanaka and Y. Higo, Phys. Rev. Lett. **87**, 026602 (2001).

[27] M. Tanaka and Y. Higo, Physica E **13**, 495 (2002).

[28] B. Grandidier, J. P. Nys, C. Delerue, D. Stiévenard, Y. Higo, and M. Tanaka, Appl. Phys. Lett. **77**, 4001 (2000).

[29] K. M. Yu, W. Walukiewicz, T. Wojtowicz, I. Kuryliszyn, X. Liu, Y. Sasaki, and J. K. Furdyna, Phys. Rev. B **65**, 201303(R) (2002).

[30] D. Chiba, F. Matsukura, and H. Ohno, Physica E **21**, 966 (2004).

[31] H. Saito, S. Yuasa, and K. Ando, Phys. Rev. Lett. **95**, 086604 (2005).

[32] M. Elsen, O. Boulle, J.-M. George, H. Jaffrès, R. Mattana, V. Cros, A. Fert, A. Lemaitre, R. Giraud, and G. Faini, Phys. Rev. B **73**, 035303 (2006).

[33] S. Ohya, I. Muneta, P. N. Hai, and M. Tanaka, Appl. Phys. Lett. **95**, 242503 (2009).

[34] D. Chiba, Y. Sato, T. Kita, F. Matsukura, and H. Ohno, Phys. Rev. Lett. **93**, 216602 (2004).

[35] A. Vedyayev, D. Bagrets, A. Bagrets, and B. Dieny, Phys. Rev. B **63**, 064429 (2001).

[36] K. Pappert, S. Hümpfner, J. Wenisch, K. Brunner, C. Gould, G. Schmidt, and L. W. Molenkamp Appl. Phys. Lett. **90**, 062109 (2007).

[37] C. Rüster, C. Gould, T. Jungwirth, J. Sinova, G. M. Schott, R. Giraud, K. Brunner, G. Schmidt, and L.W. Molenkamp, Phys. Rev. Lett. **94**, 027203 (2005).

[38] Z. Liu, J. De Boeck, V. V. Moshchalkov, and G. Borghs, J. Magn. Magn. Mater. **242**, 967 (2002).

[39] S. Ohya, P. N. Hai, Y. Mizuno, and M. Tanaka, Phys. Rev. B **75**, 155328 (2007).

[40] A. G. Petukhov, A. N. Chantis, and D. O. Demchenko, Phys. Rev. Lett. **89**, 107205 (2002).

[41] T. Hayashi, M. Tanaka, and A. Asamitsu, J. Appl. Phys. **87**, 4673 (2000).

[42] H. Shimizu and M. Tanaka, J. Appl. Phys. **91**, 7487 (2002).

[43] A. Oiwa, R. Moriya, Y. Kashimura, and H. Munekata, J. Magn. Magn. Mater. **272-276**, 2016 (2004).

[44] S. H. Chun, S. J. Potashnik, K. C. Ku, P. Schiffer, and N. Samarth, Phys. Rev. B **66**, 100408(R) (2002).

[45] R. Wessel and M. Altarelli, Phys. Rev. B **39**, 12802 (1989).

[46] J. M. Luttinger and W. Kohn, Phys. Rev. B **97**, 869 (1955).

[47] T. Dietl, H. Ohno, and F. Matsukura, Phys. Rev. B **63**, 195205 (2001).

[48] In the ideal RTD structure, the same voltage drops occur at the two barriers and no voltage drops occur in other regions (electrodes and QW), thus the ideal multiple number is 2.

[49] M. Sawicki, F. Matsukura, A. Idziaszek, T. Dietl, G. M. Schott, C. Ruester, C. Gould, G. Karczewski, G. Schmidt, and L. W. Molenkamp, Phys. Rev. B **70**, 245325 (2004).

[50] H. Shimizu, T. Hayashi, T. Nishinaga, and M. Tanaka, Appl. Phys. Lett. **74**, 398 (1999).

[51] S. Ohya, I. Muneta, P. N. Hai, and M. Tanaka, Phys. Rev. Lett. **104**, 167204 (2010).

[52] E. Likovich et al., Phys. Rev. B **80**, 201307(R) (2009).

[53] T. Niizeki, N. Tezuka, and K. Inomata, Phys. Rev. Lett. **100**, 047207 (2008).

[54] S. Ohya, K. Takata, and M. Tanaka, Nature Phys. **7**, 342 (2011).

[55] S. Ohya, I. Muneta, Y. Xin, K. Takata, and M. Tanaka, Phys. Rev. B **86**, 094418 (2012).

[56] T. Dietl, H. Ohno, F. Matsukura, J. Cibert, and D. Ferrand, Science **287**, 1019 (2000).

[57] T. Jungwirth, et al., Phys. Rev. B **72**, 165204 (2005).

[58] D. Neumaier, M. Turek, U. Wurstbauer, A. Vogl, M. Utz, W. Wegscheider, and D. Weiss, Phys. Rev. Lett. **103**, 087203 (2009).

59 Y. Nishitani, D. Chiba, M. Endo, M. Sawicki, F. Matsukura, T. Dietl, and H. Ohno, Phys. Rev. B **81**, 045208 (2010).

60 K. Hirakawa, S. Katsumoto, T. Hayashi, Y. Hashimoto, and Y. Iye, Phys. Rev. B **65**, 193312 (2002).

61 K. S. Burch, *et al.* Phys. Rev. Lett. **97**, 087208 (2006).

62 V. F. Sapega, M. Moreno, M. Ramsteiner, L. Däweritz, and K. H. Ploog, Phys. Rev. Lett. **94**, 137401 (2005).

63 K. Ando, H. Saito, K. C. Agarwal, N. C. Debnath, and V. Zayets, Phys. Rev. Lett. **100**, 067204 (2008).

64 L. P. Rokhinson *et al.* Phys. Rev. B **76**, 161201(R) (2007).

65 K. Alberi, K. M. Yu, P. R. Stone, O. D. Dubon, W. Walukiewicz, T. Wojtowicz, X. Liu, and J. K. Furdyna, Phys. Rev. B **78**, 075201 (2008).

66 J. Okabayashi, A. Kimura, O. Rader, T. Mizokawa, A. Fujimori, T. Hayashi, and M. Tanaka, Phys. Rev. B **64**, 125304 (2001).

67 H. Akai, Phys. Rev. Lett. **81**, 3002 (1998).

68 A. Richardella, P. Roushan, S. Mack, B. Zhou, D. A. Huse, D. D. Awschalom, and A. Yazdani, Science **327**, 665 (2010).

69 K. W. Edmonds *et al.* Phys. Rev. Lett. **92**, 037201 (2004).

70 K. Wagner, D. Neumaier, M. Reinwald, W. Wegscheider, and D. Weiss, Phys. Rev. Lett. **97**, 056803 (2006).

71 F. Marczinowski, J. Wiebe, J.-M. Tang, M. E. Flatté, F. Meier, M. Morgenstern, and R. Wiesendanger, Phys. Rev. Lett. **99**, 157202 (2007).

72 G. D. Sanders, Y. Sun, C. J. Stanton, G. A. Khodaparast, J. Kono, D. S. King, Y. H. Matsuda, S. Ikeda, N. Miura, A. Oiwa , and H. Munekata, Physica E **20**, 378 (2004).

73 Y. H. Matsuda, G. A. Khodaparast, M. A. Zudov, J. Kono, Y. Sun, F. V. Kyrychenko, G. D. Sanders, C. J. Stanton, N. Miura, S. Ikeda, Y. Hashimono, S. Katsumoto, and H. Munekata, Phys. Rev. B **70**, 195211 (2004).

74 K. Hirakawa, A. Oiwa, and H. Munekata, Physica E **10**, 215 (2001).

75 S. Ohya, H. Shimizu, Y. Higo, J. Sun, and M. Tanaka, Jpn. J. Appl. Phys. **41**, L24 (2002).

76 T. Slupinski, H. Munekata, and A. Oiwa, Appl. Phys. Lett. **80**, 1592 (2002)

77 S. Ohya, H. Kobayashi, and M. Tanaka, Appl. Phys. Lett. **83**, 2175 (2003).

78 A. Slobodskyy, C. Gould, T. Slobodskyy, C. R. Becker, G. Schmidt, and L. W. Molenkamp, Phys. Rev. Lett. **90**, 246601 (2003).

79 M. Tran, J. Peiro, H. Jaffrès, J.-M. George, O. Mauguin, L. Largeau, and A. Lemaître, Appl. Phys. Lett. **95**, 172101 (2009).

80 H. Ohno, N. Akiba, F. Matsukura, A. Shen, K. Ohtani, and Y. Ohno, Appl. Phys. Lett. **73**, 363 (1998).

81 M. Elsen, H. Jaffrès, R. Mattana, M. Tran, J.-M. George, A. Miard, and A. Lemaître, Phys. Rev. Lett. **99**, 127203 (2007).

82 T. Dietl and D. Sztenkiel, arXiv:1102.3267v2.

83 S. Ohya, K. Takata, I. Muneta, P. N. Hai, and M. Tanaka, arXiv:1102.4459v3.

84 T. W. Hickmott, Phys. Rev. B **46**, 15169 (1992).

85 E. E. Mendez, W. I. Wang, B. Ricco, and L. Esaki, Appl. Phys. Lett. **47**, 415 (1985).

86 M. Kobayashi, I. Muneta, Y. Takeda, Y. Harada, A. Fujimori, J. Krempasky, T. Schmitt, S. Ohya, M. Tanaka, M. Oshima, and V. N. Strocov, arXiv:1302.0063.

87 J. Fujii *et al.*, Phys. Rev. Lett. **111**, 097201 (2013).

88 M. A. Mayer, P. R. Stone, N. Miller, H. M. Smith, O. D. Dubon, E. E. Haller, K. M. Yu, W. Walukiewicz, X. Liu, and J. K. Furdyna, Phys. Rev. B **81**, 045205 (2010).

89 M. Yildirim, S. March, R. Mathew, A. Gamouras, X. Liu, M. Dobrowolska, J. K. Furdyna, and K. C. Hall, Phys. Rev. B **84**, 121202(R) (2011).

90 Q. Song, K. H. Chow, Z. Salman, H. Saadaoui, M. D. Hossain, R. F. Kiefl, G. D. Morris, C. D. P. Levy, M. R. Pearson, T. J. Parolin, I. Fan, T. A. Keeler, M. Smadella, D. Wang, K. M. Yu, X. Liu, J. K. Furdyna, and W. A. MacFarlane, Phys. Rev. B **84**, 054414 (2011).

91 B. C. Chapler, R. C. Myers, S. Mack, A. Frenzel, B. C. Pursley, K. S. Burch, E. J. Singley, A. M. Dattelbaum, N. Samarth, D. D. Awschalom, and D. N. Basov, Phys. Rev. B **84**, 081203(R) (2011).

92 M. Dobrowolska, K. Tivakornsasithorn, X. Liu, J. K. Furdyna, M. Berciu, K. M. Yu, and W.Walukiewicz, Nature Mater. **11**, 444 (2012).

[93] B. C. Chapler, S. Mack, R. C. Myers, A. Frenzel, B. C. Pursley, K. S. Burch, A. M. Dattelbaum, N. Samarth, D. D. Awschalom, and D. N. Basov, Phys. Rev. B **87**, 205314 (2013).

[94] R. Bouzerar and G. Bouzerar, Europhys. Lett. **92**, 47006 (2010).

[95] I. Muneta, H. Terada, S. Ohya, and M. Tanaka, Appl. Phys. Lett. **103**, 032411 (2013).

[96] T. Jungwirth, Jairo Sinova, A. H. MacDonald, B. L. Gallagher, V. Novák, K. W. Edmonds, A. W. Rushforth, R. P. Campion, C. T. Foxon, L. Eaves, E. Olejník, J. Mašek, S.-R. Eric Yang, J. Wunderlich, C. Gould, L. W. Molenkamp, T. Dietl, and H. Ohno, Phys. Rev. B **76**, 125206 (2007).

[97] T. Jungwirth, P. Horodyská, N. Tesařová, P. Němec, J. Šubrt, P. Malý, P. Kužel, C. Kadlec, J. Mašek, I. Němec, M. Orlita, V. Novák, K. Olejník, Z. Šobáň, P. Vašek, P. Svoboda, and Jairo Sinova, Phys. Rev. Lett. **105**, 227201 (2010).

[98] J. Mašek, F. Máca, J. Kudrnovský, O. Makarovsky, L. Eaves, R. P. Campion, K. W. Edmonds, A. W. Rushforth, C. T. Foxon, B. L. Gallagher, V. Novák, Jairo Sinova, and T. Jungwirth, Phys. Rev. Lett. **105**, 227202 (2010).

[99] K. Alberi, J. Wu, W. Walukiewicz, K. M. Yu, O. D. Dubon, S. P. Watkins, C. X. Wang, X. Liu, Y.-J. Cho, and J. Furdyna, Phys. Rev. B **75**, 045203 (2007).

[100] J. S. Blakemore, Winfield J. Brown Jr, Merrill L. Stass, and Dustin A. Woodbury, J. Appl. Phys. **44**, 3352 (1973).

[101] Dale. E. Hill, J. Appl. Phys. **41**, 1815 (1970).

[102] P. Mahadevan and A. Zunger, Appl. Phys. Lett. **85**, 2860 (2004).

[103] J.-M. Tang and M. E. Flatté, Phys. Rev. Lett. **92**, 047201 (2004).

[104] Y. Mizuno, S. Ohya, P. N. Hai, and M. Tanaka, Appl. Phys. Lett. **90**, 162505 (2007).

[105] K. Mizushima, T. Kinno, T. Yamauchi, and K. Tanaka, IEEE Trans. Magn. **33**, 3500 (1997).

[106] S. van Dijken, X. Jiang, and S. S. P. Parkin, Appl. Phys. Lett. **83**, 951 (2003).

[107] Y. W. Huang, C. K. Lo, Y. D. Yao, L. C. Hsieh, and D. R. Huang, IEEE Trans. Magn. **41**, 2682 (2005).

[108] S. Ohya, I. Muneta, and M. Tanaka, Appl. Phys. Lett. **96**, 052505 (2010).

[109] M. I. Lepsa *et al.*, 1997 International Semiconductor Conference 20th Edition. CAS'97 Proceedings (Cat. No. 97TH8261) **1**, 139 (1997).

[110] C. Ertler and J. Fabian, Phys. Rev. B **75**, 195323 (2007).

[111] I. Muneta, S. Ohya, and M. Tanaka, Appl. Phys. Lett. **100**, 162409 (2012).

[112] T. Uemura, T. Marukame, and M. Yamamoto, IEEE Trans. Magn. **39**, 2809 (2003).

[113] C. Ertler, W. Pötz, and J. Fabian, Appl. Phys. Lett. **97**, 042104 (2010).

[114] T. Uemura, S. Honma, T. Marukame, and M. Yamamoto, Jpn. J. Appl. Phys. **43**, L44 (2004).

[115] H. Ohno, D. Chiba, F. Matsukura, T. Omiya, E. Abe, T. Dietl, Y. Ohno, and K. Ohtani, Nature **408**, 944 (2000).

[116] S. Koshihara, A. Oiwa, M. Hirasawa, S. Katsumoto, Y. Iye, C. Urano, H. Takagi, and H. Munekata, Phys. Rev. Lett. **78**, 4617 (1997).

[117] M. Tanaka, H. Shimizu, and T. Hayashi, J. Vac. Sci. Technol. A **18**, 1247 (2000).

[118] H. Ohno, J. Magn. Magn. Mater. **200**, 110 (1999).

[119] M. Tanaka, J. Vac. Sci. Technol. B **16**, 2267 (1998).

[120] P. N. Hai, L. D. Anh, S. Mohan, T. Tamegai, M. Kodzuka, T. Ohkubo, K. Hono, and M. Tanaka, Appl. Phys. Lett. **101**, 182403 (2012), and Supplementary Material.

[121] P. N. Hai, L. D. Anh, and M. Tanaka, Appl. Phys. Lett. **101**, 252410 (2012), and Supplementary Material.

[122] M. Kodzuka, T. Ohkubo, and K. Hono, Ultramicroscopy **111**, 557 (2011).

[123] E. Malguth, A. Hoffmann, and M. R. Phillips, Phys. Stat. Sol. (b) **245**, 455 (2008).

[124] T. L. Estle, Phys. Rev. **136**, A1702 (1964).

[125] S. Haneda, M. Yamaura, Y. Takatani, K. Hara, S. Harigae, H. Munekata, Jpn. J. Appl. Phys. **39**, L9 (2000).

[126] M. Takushima, Y. Kajikawa, Phys. Stat. Sol. (c) **5**, 2781 (2008).

[127] S. Haneda, in Binary Alloy Phase Diagrams, edited by T. B. Massalski (ASM International, Ohio, 1990), p. 279.

[128] K. Ando, H. Munekata, J. Magn. Magn. Mat. **272-276**, 2004 (2004).

[129] Recently, ferromagnetism was reported in

n-type Co-doped $TiO_2$ (Ref.130). However, the intrinsic ferromagnetism in Co-doped $TiO_2$ is controversial, because the MCD spectrum of Co-doped $TiO_2$ does not show enhancement at optical critical point energies of $TiO_2$, while it is enhanced at energies not related to the band structure of $TiO_2$, and very broad MCD signals are seen at energies smaller than the band gap of $TiO_2$ (Ref.131).

[130] Y. Yamada, K. Ueno, T. Fukumura, H. T. Yuan, H. Shimotani, Y. Iwasa, L. Gu, S. Tsukimoto, Y. Ikuhara, and M. Kawasaki, Science **332**, 1065 (2011).

[131] T. Fukumura, Y. Yamada, K. Tamura, K. Nakajima, T. Aoyama, A. Tsukazaki, M. Sumiya, S. Fuke, Y. Segawa, To. Chikyow, T. Hasegawa, H. Koinuma, and M. Kawasaki, Jpn. J. Appl. Phys. **42**, L105 (2003).

[132] H. Okamoto, J. Phase Equil. **12**, 457 (1991).

[133] A. K. L. Fan, G. H. Rosenthal, H. L. Mckinzie, and A. Wold, J. Solid State Chem. **5**,136 (1972).

[134] P. N. Hai, D. Sasaki, L. D. Anh, and M. Tanaka, Appl. Phys. Lett. **100**, 262409 (2012).

[135] K. Sato, H. Katayama-Yoshida, and P. H. Dederichs, Jpn. J. Appl. Phys. **44**, L948 (2005).

[136] P. N. Hai, S. Yada, and M. Tanaka, J. Appl. Phys. **109**, 073919 (2011).

[137] M. Yokoyama, H. Yamaguchi, T. Ogawa, and M. Tanaka, J. Appl. Phys. **97**, 10D317 (2005).

[138] P. N. Hai, S. Ohya, M. Tanaka, S. E. Barnes, and S. Maekawa, Nature **458**, 489 (2009).

[139] S. Kuroda, N. Nishizawa, K. Takita, M. Mitome, Y. Bando, K. Osuch and T. Dietl, Nature Mat. **6**, 440 (2007).

[140] Y. Shuto, M. Tanaka, and S. Sugahara, Appl. Phys. Lett. **90**, 132512 (2007).

[141] A. Soibel, E. Zeldov, M. Rappaport, Y. Myasoedov, T. Tamegai, S. Ooi, M. Konczykowski, V. B. Geshkenbein, Nature **406**, 282 (2000).

[141] B. Lee, T. Jungwirth, and A. H. MacDonald, Semicond. Sci. Technol. **17**, 393 (2002).

[142] B. Lee, T. Jungwirth, and A. H. MacDonald, Semicond. Sci. Technol. **17**, 393 (2002).

[143] I. Zutic, J. Fabian, and S. Das Sarma, Phys. Rev. Lett. **88**, 066603 (2002).

[144] J. Fabian, I. Zutic, S. Das Sarma, Phys. Rev. B **66**, 165301 (2002).

[145] N. Lebedeva and P. Kuivalainen, J. Appl. Phys. **93**, 9845 (2003).

[146] J. Fabian, I. Zutic, Phys. Rev. B **69**, 115314 (2004).

[147] M. E. Flatte, Z. G. Yu, E. Johnston-Halperin and D. D Awschalom, Appl. Phys. Lett. **82**, 4740 (2003).

[148] S. Sugahara and M. Tanaka, J. Appl. Phys. **97**, 10D503 (2005).

[149] T. Matsuno, S. Sugahara, and M. Tanaka, Jpn. J. Appl. Phys. **43**, 6032 (2004).

[150] M. Sawicki, D. Chiba, A. Korbecka, Y. Nishitani, J. A. Majewski, F. Matsukura, T. Dietl, H. Ohno, Nature Phys. **6**, 22 (2010).

[151] N. A. Semikolenova, I. M. Nesmelova, E. N. Khabarov, Sov. Phys. Semicond. **12**, 1139 (1978).

[152] M. Kobayashi, P. N. Hai, L. D. Anh, T. Schmitt, A. Fujimori, M. Tanaka, M. Oshima, and V. N. Strocov, unpublished.

[153] J. Okabayashi, A. Kimura, A. Fujimori, T. Hayashi, and M. Tanaka, Phys Rev. B **58**, R4211 (1998).

[154] P. W. Anderson, Phys. Rev **124**, 41 (1964).

[155] J. R. Schrieffer and P. A. Wolff, Phys. Rev. **149**, 491 (1966).

[156] R. E. Walstedt and L. R. Walker, Phys. Rev. **B11**, 3280 (1975).